\documentclass[reprint,amsmath,amssymb,aps,prc,nofootinbib]{revtex4-1}
\usepackage{bm}% bold math
\usepackage{graphicx}
\usepackage{dcolumn}% Align table columns on decimal point
\usepackage{color}
\newcommand{\be}{\begin{equation}}
\newcommand{\ee}{\end{equation}}
\newcommand{\bea}{\begin{eqnarray}}
\newcommand{\eea}{\end{eqnarray}}

\newcommand{\mbss}[1]{_{\mbox{\scriptsize #1}}}

\newcommand{\vphu}{\vphantom{*}}

\newcommand{\ve}{\varepsilon}

\begin{document}

\title{Multiphonon structure of high-spin states
in $^{40}$Ca, $^{90}$Zr, and $^{208}$Pb
}
\date{\today}
\author{N. Lyutorovich}
\affiliation{St. Petersburg State University, St. Petersburg, 199034,
  Russia}
\author{V. Tselyaev}
%\email{tselyaev@mail.ru}
\affiliation{St. Petersburg State University, St. Petersburg, 199034,
  Russia}

\date{\today}

\begin{abstract}
The method of description of the high-spin states, which was previously
developed and applied for the states of this type in $^{208}$Pb,
is generalized for the case of the states having more complex multiphonon structure.
In this method,
the harmonic approximation with the renormalized phonons is used in which
the phonons themselves are determined within the non-linear version
of the model based on the random-phase approximation (RPA)
and including both the RPA correlations and the beyond-RPA ones.
The mean field and the residual interaction are derived within the framework of
the self-consistent RPA from the energy-density functional of the Skyrme type.
The method is applied for the analysis of the
available experimental data in doubly magic $^{208}$Pb and $^{40}$Ca and
in semi-magic $^{90}$Zr.
\end{abstract}

\maketitle

%---------------------
\section{Introduction}
%---------------------

High-spin states in nuclei have been the subject of the experimental
and theoretical investigations for a long time
(see reviews~\cite{VoigtRMP83,SaladinBook91,Ward2001})
and continue to be a topic of current interest, see, e.g., Refs.
\cite{Meng_2016,Frauendorf_2018, Petrache_2019,Kumar_2020,Afanasjev_2022,Yoshida_2022}.
The most part of the data on the high-spin states refers to the non-magic nuclei,
however the recent experiments and theoretical analysis show that
there are also long series of high-spin states in magic and
semi-magic nuclei
\cite{Rudolph_1999,Ideguchi_2001,Inakura_2002,Caurier_2007,Chiba_2014,
Broda_2017,Sakai_2020,Wang_2020,Dey_2022}.
At present, the high-spin states
have been experimentally identified only in two doubly-magic nuclei,
$^{208}$Pb and $^{40}$Ca, and in the semi-magic nucleus $^{90}$Zr.
The first observation of high-spin states
in these nuclei has been made
in $^{90}$Zr in Ref.~\cite{Warburton_1985}
where the states up to spin $I=20$
were investigated and
the shell-model (SM) calculations were performed.
The large-scale SM calculations for rotational bands in $^{90}$Zr
were also presented in Refs.~\cite{Wang_2021,Dey_2022}
where previous experiments and calculations are also reviewed.
At the same time, the close coexistence of spherical states
having n-particle--n-hole (npnh) configuration structure
with deformed and superdeformed rotational bands in
magic and semi-magic nuclei significantly complicates
their theoretical description.

Experimental and theoretical studies of $^{40}$Ca nucleus show a very complex
structure of its excited states, where the spherical states and
approximately five rotational bands, including normally deformed (ND)
and superdeformed (SD) bands,
exist~\cite{Ideguchi_2001,Caurier_2007,Chen_2017,Sakai_2020}.
The studies show that the first and the second $0^+$ excited
states in this nucleus, i.e. ND and SD band-heads,
have 4p4h and 8p8h configuration structures, respectively.
Calculations for the ND and SD bands were performed
in the cranked relativistic mean-field~\cite{Ideguchi_2001}
and cranked HF or HFB \cite{Inakura_2002,Oi_2007,Sakai_2020} models.
The SD-band properties were also studied in the framework
of the cranking covariant density functional theory
with pairing in the shell-model-like approach with conserved
particle numbers~\cite{Wang_2020}.
The transition (not band-head) energies and quadrupole moments
of the ND and SD bands were well reproduced in the large-scale SM calculations
in Ref.~\cite{Caurier_2007}.
It should be noted that, in all these methods, calculations
overestimate the energy of the ND and SD band-head states.
However, besides the ND and SD bands, there are many other states in
$^{40}$Ca which have
not yet been described within the microscopic approach.

The doubly magic $^{208}$Pb is of particular interest
since many high-spin states have been experimentally assigned in this nucleus
and since it is a conventional laboratory to test the theory.
Very important new experimental data for $^{208}$Pb
up to spin $I=30$ together with a theoretical analysis
in the framework of the
SM have been presented
in Ref.~\cite{Broda_2017}.
The advantage of the SM calculations is that they allow one to take
into account many complex configurations.
However, the model of Ref.~\cite{Broda_2017}
is not self-consistent in particular because
the single-particle (sp) energies used in the calculations
were adjusted to reproduce
the experimental spectra of the neighboring odd nuclei.

The self-consistent description of the high-spin states in $^{208}$Pb
has been presented in Ref.~\cite{Lyutorovich_2022} within
the renormalized time-blocking approximation (RenTBA)
developed in Ref.~\cite{Tselyaev_2018}.
The RenTBA is the non-linear version
of the model based on the random-phase approximation (RPA)
and including both the RPA correlations and the beyond-RPA ones.
The main configuration space of the RenTBA consists of
the 1p1h and $\text{1p1h}\otimes\text{phonon}$ configurations
where the phonons are determined self-consistently from the
non-linear RenTBA equations and thus include correlations beyond the RPA.
The more complex configurations are included in the RenTBA in part due to the
ground-state correlations (of the RPA type) and the non-linear effects
(see \cite{Tselyaev_2018} for more details).
In Ref.~\cite{Lyutorovich_2022} it was obtained that the uncoupled
$\text{1p1h}\otimes\text{phonon}$ configurations are rather good
approximation for the high-spin states in $^{208}$Pb.
However, in this model the correlations between the 1p1h components
of the $\text{1p1h}\otimes\text{phonon}$ configurations are neglected.
The inclusion of these correlations means the replacement of
the $\text{1p1h}\otimes\text{phonon}$ configurations with
the $\text{phonon}\otimes\text{phonon}$ ones.
The results of Ref.~\cite{Lyutorovich_2022} show that the effect
of this replacement can be noticeable.
One can expect that this effect will increase at the increase of
the spin of the state resulting in the increase of its configuration
complexity.

The most elaborated approach in which the multiphonon configurations
are included explicitly is the quasiparticle-phonon model (QPM)
of Soloviev and co-workers (see
\cite{Vdovin_1983,Voronov_1983,Soloviev_1992,Bertulani_1999}
and references therein and also Refs. \cite{Giai_1998,Severyukhin_2018prc}
where the self-consistent version of this model was developed).
In the QPM, the interaction between the phonons is taken into account,
but the phonons are determined as the solutions of the RPA or
the quasiparticle RPA equations.
More recently, the multiphonon models were developed in
Refs. \cite{Andreozzi_2007,Bianco_2012,De_Gregorio_2016}
within the framework of an equation of motion phonon method (EMPM),
in which the phonons are introduced in the Tamm-Dancoff approximation,
and in Ref.~\cite{Litvinova_2015} within the covariant nuclear response theory
based on the relativistic quasiparticle time-blocking approximation.
However, to our knowledge, these models including QPM were not applied
to the study of the multiphonon high-spin states.

Consistent multiphonon model should certainly takes into account
interaction between the phonons and restrictions imposed by the Pauli principle.
The phonon-phonon interaction is especially important in the models
in which the phonon basis is fixed as it takes place, e.g., in the QPM and EMPM.
In this case the phonon-phonon interaction can affect the properties of
both multiphonon and one-phonon states, in particular, their energies.
In the method used in Ref.~\cite{Lyutorovich_2022}, the effect of the phonon-phonon
interaction on the one-phonon states is incorporated by means of the
phonon renormalization within the RenTBA.
The effect of this interaction on the two-phonon high-spin states
(composed of the {\it renormalized} phonons)
was found to be small by comparing the energies
of the pure $\text{1p1h}\otimes\text{phonon}$ configurations with the results
of the full-scale RenTBA calculations.

The goal of the present paper is to extend the method of Ref.~\cite{Lyutorovich_2022}
to the high-spin states having multiphonon structure.
The calculation scheme is fully self-consistent and is based on
the energy-density functional (EDF) of the Skyrme type.
Calculations are performed for the high-spin states in $^{40}$Ca,
$^{90}$Zr, and $^{208}$Pb.
The results are compared with available experimental data.

%-------------------------------
\section{Theoretical framework and calculation scheme} \label{sec:theor}
%------------------------------------------

Our approach is based on
the harmonic approximation with the renormalized phonons.
Thus, the basic elements of the theory are the phonons which are determined within the RenTBA.
The RenTBA (see \cite{Tselyaev_2018} for more details) is a non-linear version of
the time-blocking approximation which is a model of the extended RPA type
including $\text{1p1h}\otimes\text{phonon}$ configurations on top of the
configurations incorporated in the RPA. The renormalized phonons are described by
the solutions of the (non-linear) RenTBA equations and therefore include 1p1h,
$\text{1p1h}\otimes\text{phonon}$, and more complex configurations resulting from
the non-linear effects. As was shown in \cite{Tselyaev_2018,Tselyaev_2020},
the renormalization reduces the energies of the phonons as compared to their RPA values
that usually decreases discrepancy with the experiment in the case of the self-consistent
scheme based on the Skyrme EDFs,
see Tables \ref{table:208Pb_phonons}--\ref{table:90Zr_phonons} in the Appendix.
The harmonic approximation means that the interaction between 1p1h and
more complex configurations is taken into account in the renormalized phonons
(having relatively low energies)
but not between them in the multiphonon configurations at higher energies
and thus the energy of the multiphonon state is determined
as a simple sum of the energies of the renormalized phonons.
In what follows, we call this approach the multiphonon model in the harmonic approximation,
or, for brevity, the multiphonon model.
To minimize violation of the Pauli principle in the multiphonon configurations
we use a simple method in which incompatible phonon combinations are excluded
with the help of the numerical analysis of the 1p1h structure of the main (RPA)
components of the phonons.

The technical details of the calculations are the following.
The single-particle (s.p.) basis and residual interaction
were calculated by the variational method for the Skyrme EDF
as described in Ref. \cite{Lyutorovich_2016}.
At the first step, the phonons are calculated in the RPA and,
at the second step, they are self-consistently optimized
in the RenTBA that implies the solution of the system
of non-linear equations ~\cite{Tselyaev_2018}.

In the calculations, the Skyrme EDF with
the parameter set SV-bas$_{-0.44}$ was used.
This set were obtained in Ref. \cite{Tselyaev_2020}
on the base of the parametrization SV-bas \cite{Kluepfel_2008} to reproduce
the basic experimental characteristics of the $M1$ excitations in $^{208}$Pb
within the RenTBA and at the same time to describe
the nuclear ground-state properties with approximately the
same accuracy as the original SV-bas set.

Wave functions and fields were represented on a spherical grid in
coordinate space.  The s.p. basis was discretized by imposing box
boundary condition with a box radius equal to 18~fm.
The particle's energies $\ve^{\vphu}_{p}$ were limited
by the maximum value $\ve^{\mbss{max}}_{p} = 100$ MeV.
The details of solving the non-linear RenTBA equations are
described in Ref.~\cite{Tselyaev_2020}.

\section{Results and discussion}\label{sec:results}

The high-spin states are in the energy region where the level density
is high, so there may be uncertainties in comparing the theoretical
values with the data. The uncertainties are not very essential
in the given investigation because we consider only low levels
for every spin and parity.
For $^{208}$Pb and $^{90}$Zr, we study yrast, yrare, and those
additional levels for which experimental data are available.
For $^{40}$Ca, where both spherical and deformed states are known,
we investigate all the states excluding the ND and SD rotational bands
because the method can not describe the states having a large deformation.
At the same time, for $^{40}$Ca and $^{90}$Zr, we also considered states
with low spins in order to give a complete picture of the states in these nuclei.
The number of known low-spin states in $^{208}$Pb is very large,
so they deserve a separate study
which goes beyond the scope of this article.

%--------------------------------
\subsection{$^{208}$Pb}%\nonumber
%--------------------------------

The results for the high-spin states in $^{208}$Pb and
a comparison with available data are presented
in Tables~\ref{table:208Pb_positive_parity} (positive parity states)
and \ref{table:208Pb_negative_parity} (negative parity states),
where $I$ and $\pi$ denote spin and parity of a state, $n$
is a number of the level for the given $I^\pi$, $E$ are
experimental and theoretical energies (in MeV).
The experimental values were taken from Refs.~\cite{NDS_2007,Broda_2017}.
The letters next to the theoretical values denote the phonon
configuration for every multiphonon state.
The correspondence of these letters to phonons and the structure
of the RPA phonons are shown in the Appendix, Table~\ref{table:208Pb_phonons}.
The renormalized phonons have more complex form where the RPA state
presents the main part and, in addition, there are many additional
components. Thus, the RPA phonon structure reflects the most significant
part of the renormalized phonons.
The renormalization of phonons changes the phonon energies of $^{208}$Pb
by 0.3 -- 1.4 MeV, see Appendix, Table~\ref{table:208Pb_phonons},
and the renormalization is especially large for high-spin phonons.

In some cases, the spin-parity assignment of levels does not match
in Refs.~\cite{Broda_2017} and \cite{NDS_2007}, namely, for the levels
of 7.974, 8.027, and 9.061 MeV.  In these cases, we preferred
the \cite{Broda_2017} identification, because it is more recent
and more substantiated. Our calculations confirm the correctness
of this choice.
\begin{table}[h]
\caption{\label{table:208Pb_positive_parity}
Positive parity states in $^{208}$Pb obtained in
the multiphonon
model with renormalized phonons for the SV-bas$_{-0.44}$
Skyrme parametrization.
Here, $I^\pi$ denotes spin and parity of a level, $n$ is a number of
the level for the given $I^\pi$, $E$ are
experimental \cite{NDS_2007,Broda_2017} and theoretical energies.
The letters next to the theoretical values denote phonons: see Appendix,
Table~\ref{table:208Pb_phonons}.}
\begin{ruledtabular}
\begin{tabular}{ccl|ccl}
$I^\pi_n$ & \multicolumn{2}{c|}{$E$[MeV]} & $I^\pi_n$ & \multicolumn{2}{c}{$E$[MeV]} \\[\smallskipamount]
         & Exp.    & SV-bas$_{-0.44}$ &  & Exp.  & SV-bas$_{-0.44}$ \\
\hline
\vspace{-2mm}
         &         &          &          &         &         \\
$14^+_1$ &         & 9.16 bc  & $13^+_1$ &         & 9.00 bc \\
$14^+_2$ &         & 9.36 cc  & $13^+_2$ &         & 9.21 dj \\
$16^+_1$ &         & 9.21 bc  & $15^+_1$ &  \footnotemark[1]
                                                   & 9.21 bc \\
$16^+_2$ &         & 9.35 cc  & $15^+_2$ &         & 9.33 cc \\
$18^+_1$ & 9.1030  & 9.44 cc  & $17^+_1$ & 9.061 \footnotemark[1]
                                                   & 9.43 cc \\
$18^+_2$ &         & 9.82 ck  & $17^+_2$ &         & 9.68 ar \\
$20^+_1$ & 10.1959 & 9.51 cc  & $19^+_1$ & 9.394   & 9.48 cc \\
$(20^+_2)$& 10.3573& 10.19 ck & $(19^+)$ \footnotemark[2]
                                         & 10.1358 & 10.04 ck \\
$20^+_3$ & 10.3710 & 10.63 cl & $19^+$   & 10.1362 & 10.17 cu \\
$(20^+_4)$& 10.5313& 10.64 cm & $21^+_1$ &         & 9.50 cc \\
$(20^+_5)$& 10.5524& 10.95 kl & $21^+_2$ &         & 10.38 ck \\
$22^+_1$ &         & 10.96 cl & $23^+_1$ & 11.3609 & 12.17 ll \\
$22^+_2$ &         & 11.60 ll & $23^+_2$ &         & 12.90 pr \\
$24^+_1$ & 11.9582 & 12.96 rr & $25^+_1$ & 12.9493 & 13.14 qr \\
$24^+_2$ &         & 13.00 qr & $25^+_2$ &         & 13.29 rr \\
$26^+_1$ &         & 13.11 rr & $27^+_1$ &         & 13.28 kk \\
$26^+_2$ &         & 13.30 qr & $27^+_2$ &         & 14.06 ccc \\
$28^+_1$ &         & 14.20 ccc& $29^+_1$ &         & 14.23 ccc \\
$28^+_2$ &         & 14.61 cck& $29^+_2$ &         & 14.60 cck \\
$30^+_1$ &         & 14.24 ccc&          &         &       \\
$30^+_2$ &         & 14.89 ccg&          &         &       \\
\end{tabular}
\end{ruledtabular}
\footnotetext[1]{The spin-parity assignment to the level 9.061
is $(17^+)$ in the NDS~\cite{NDS_2007} and $17^+$
in Ref.~\cite{Broda_2017} but $15^+$ in Ref.~\cite{Heusler_2020}.}
\footnotetext[2]{The spin-parity assignment to the level 10.1358
in Ref.~\cite{Broda_2017} is $(18^-,19^+)$.}
\end{table}
\begin{table}[h]
\caption{\label{table:208Pb_negative_parity}
The same as in Table \ref{table:208Pb_positive_parity} but
for the negative parity states in $^{208}$Pb.}
\begin{ruledtabular}
\begin{tabular}{ccl|ccl}
$I^\pi_n$ & \multicolumn{2}{c|}{$E$[MeV]} & $I^\pi_n$ & \multicolumn{2}{c}{$E$[MeV]} \\[\smallskipamount]
         & Exp.    & SV-bas$_{-0.44}$  &  & Exp. & SV-bas$_{-0.44}$ \\
\hline
\vspace{-2mm}
         &         &          &          &         &       \\
$13^-_1$ & 6.448   & 6.47 r   & $14^-_1$ & 6.743   & 6.81 r \\
$13^-_2$ & 7.5288  & 7.62 ac  & $14^-_2$ & 7.974 \footnotemark[2]
                                                 & 7.72 ac \\
$15^-_1$ & 8.027 \footnotemark[1]
                   & 8.34 cd  & $16^-_1$ & 8.562   & 8.70 ce \\
$(15^-_2)$& 8.1506 & 8.66 ce  & $16^-_2$ & 8.6007  & 9.03 cd \\
$15^-_3$ & 8.2645  & 8.86 cf  & $16^-_3$ & 8.7235  & 9.06 cf \\
$15^-_4$ & 8.3508  & 9.04 dk  & $18^-_1$ & 10.1358 \footnotemark[3]
                                                   & 10.29 ci \\
$17^-_1$ & 8.8128  & 9.36 ch  & $18^-_2$ &         & 10.35 cj \\
$17^-_2$ &         & 9.57 ct  & $20^-_1$ & 10.3419 & 11.07 cp \\
$19^-_1$ &         & 10.55 cj & $20^-_2$ &         & 11.13 cr \\
$19^-_2$ &         & 11.03 br & $22^-_1$ &         & 11.22 cr \\
$21^-_1$ & 10.9343 & 11.15 cr & $22^-_2$ &         & 11.35 cq \\
$21^-_2$ &         & 11.22 co & $24^-_1$ &         & 11.47 cr \\
$23^-_1$ &         & 11.40 cr & $24^-_2$ &         & 12.28 lr \\
$23^-_2$ &         & 11.77 cs & $26^-_1$ & 13.5360 & 13.35 cce \\
$25^-_1$ &         & 12.87 lr & $26^-_2$ &         & 13.88 ccg \\
$25^-_2$ &         & 13.30 cce& $28^-_1$ & 13.6747 & 14.93 cci \\
$27^-_1$ &         & 14.08 ccg& $28^-_2$ &         & 15.00 ccp \\
$27^-_2$ &         & 14.22 cct& $30^-_1$ &         & 15.72 ccp \\
$29^-_1$ &         & 15.20 ccj& $30^-_2$ &         & 15.78 ccr \\
$29^-_2$ &         & 15.68 bcr&          &         &          \\

\end{tabular}
\end{ruledtabular}
\footnotetext[1]{The NDS \cite{NDS_2007} assignment to the level 8.027 is $(14^-)$
but Ref.~\cite{Broda_2017} defines this level as $15^-$.}
\footnotetext[2]{The NDS \cite{NDS_2007} spin-parity assignment to the level
7.974 is $(15^-)$ but \cite{Broda_2017} assigns this level as $14^-$.}
\footnotetext[3]{The spin-parity assignment to the level 10.1358
in Ref.~\cite{Broda_2017} is $(18^-,19^+)$.}
\end{table}

Tables~\ref{table:208Pb_positive_parity} and \ref{table:208Pb_negative_parity}
show that results of the self-consistent calculations for $^{208}$Pb
in the framework of the
multiphonon
model are in
fairly well
agreement with the experimental data without any
refit of the parameters.
The close results were obtained in Ref.~\cite{Lyutorovich_2022} with
parametrization SKXm$_{-0.49}$ \cite{Tselyaev_2020},
but here we extend the results to the three-phonon states.
An important point is that, while there are many close
two- and three-phonon states, the yrast and yrare nuclear levels
are described just as the lowest states for the given spin and parity.
Of course, the model cannot establish a one-to-one correspondence
between theoretical and experimental values for these states.
But for the yrast and yrare levels, the energies are predicted
with an accuracy no worse than the difference between the energies
of these levels.

\begin{figure}[h]
\includegraphics[scale=0.37]{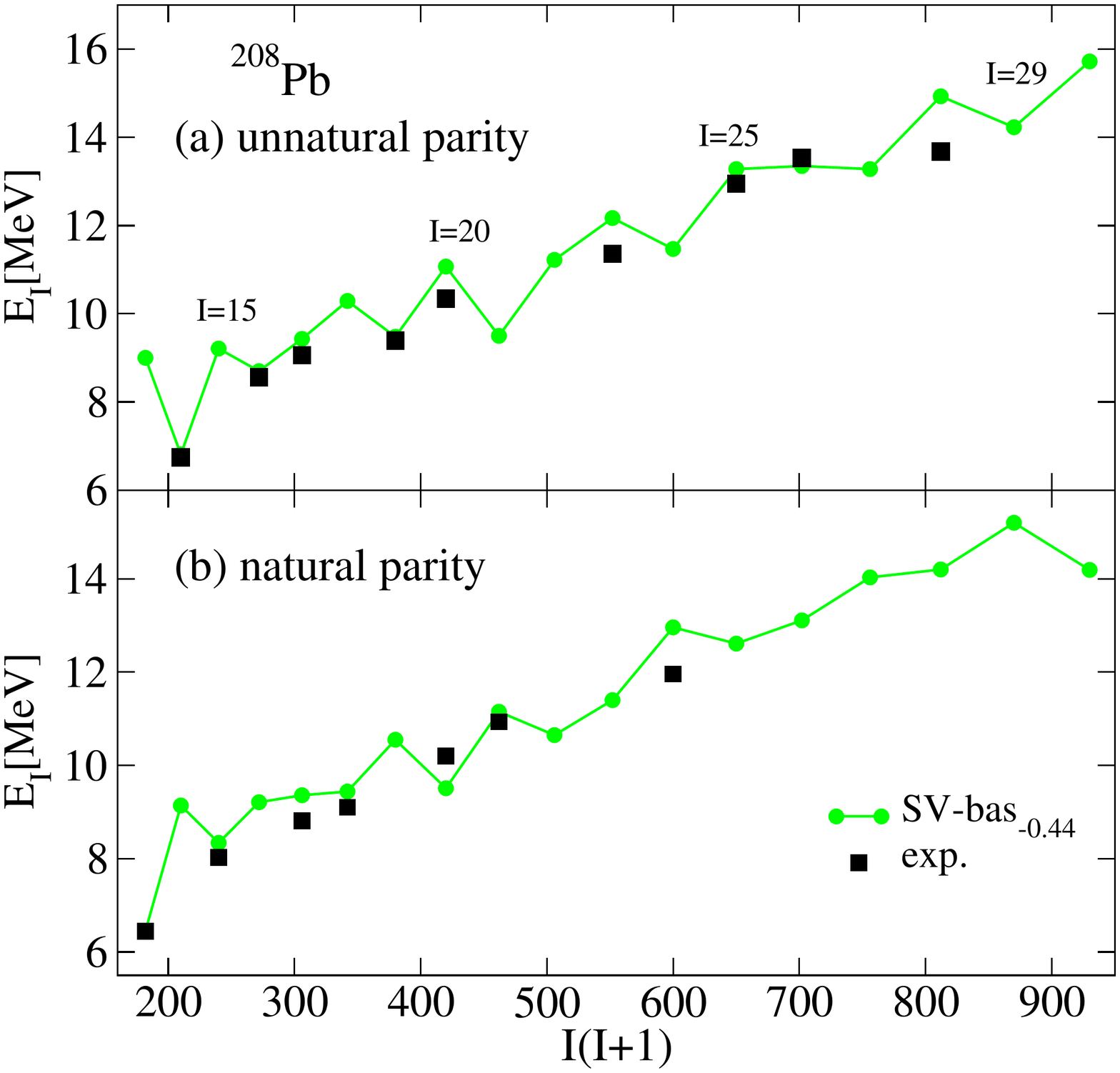}
\caption{\label{fig:208Pb_yrast_bands} The energies $E_1$ of $^{208}$Pb yrast
states calculated with RenTBA multiphonon model
for the SV-bas$_{-0.44}$ Skyrme parametrization (green line with circles)
are shown as function of $I(I+1)$ for the spins $ 13 \leqslant I \leqslant 30$
and compared with available experimental data \cite{NDS_2007,Broda_2017}
(black squares).
}
\end{figure}

The energies of the yrast states in $^{208}$Pb
as a function of $I(I+1)$ in the spin range $13 \le I \le 30$
are presented in Fig.~\ref{fig:208Pb_yrast_bands}.
Here, the energies $E$ (in MeV) calculated in the
multiphonon
model with the SV-bas$_{-0.44}$ functional are shown by green dots
connected by lines of the same color, the experimental data
\cite{NDS_2007,Broda_2017} are given in black squares.

Our new results confirm the conclusions made in the previous
paper, Ref.~\cite{Lyutorovich_2022}.
The general trend looks very much like a rotational band,
though the experimental trend in detail does not always
follow exactly a straight line: $E$ is approximately constant
in the ranges $I = 17-18$ and $26-28$.
This deviation is easy explained in the framework of the
multiphonon
model. These parts with the constant $E$ values arise as parts
of the phonon multiplets with some additional details
arisen because of the phonon renormalization.
Nevertheless, on the whole,
both experimental and theoretical
trends are similar to a rotational band.
However, this band does not rely on a collective rotation
that is hindered by the spherical shape and the large shell
gap in $^{208}$Pb. The analysis of the microscopic structure
shows that the high-spin states in $^{208}$Pb have predominantly
simple form in terms of s.p. excitations
since these states have a simple form in terms of phonons
(i.e., they are two- and three phonon states)
and, at the same time, the phonons are approximately
1p1h states (see the in Table~\ref{table:208Pb_phonons}).
Some complication arises because of the renormalization
of the phonons, but this is just the same renormalization
that changes the bare phonons appearing in the self-consistent RPA
to the dressed phonons appearing as the $^{208}$Pb
lower spin states observed in the experiment.
The rotational trend stems from a change in the angular momentum
of one single nucleon. Here the rotational part of the
s.p. kinetic energy gives a large contribution to the nucleon energy
and thereby to the change of nucleus energy.

The calculations show the following structure of the high-spin states
in $^{208}$Pb. The $13^-_1$ and $14^-_1$
states have one-phonon configurations
while the other states with spins $13\leqslant I \leqslant 26$
including $13^-_2$ and $14^-_2$ but except for $26^-$
are two-phonon configurations.
The $26^+_2$ and $27^+_1$ states are of the two-phonon
nature while all the other states with $I = 27$ and
the even larger $I$ need
the treatment in terms of the
three-phonon configurations.
Of course, we described here only several lowest states
for every spin and parity. For higher energies,
%the two-, three-, and other n-phonon configurations
%have close energies and
the structure of states
becomes more complicated.

%------------------------------
\subsection{$^{40}$Ca}%\nonumber
%------------------------------

As it was mentioned, we do not consider the ND and SD rotational
bands in $^{40}$Ca based on the $0^+_2$ and $0^+_3$ states.
The ND and SD band-head states have,
in the deformed basis,
the 4p4h and 8p8h configuration structure,
respectively, and it is very difficult
to describe such states in the spherical basis.
The results for all other states in $^{40}$Ca are presented
in Tables~\ref{table:40Ca_positive_parity} (positive parity states)
and \ref{table:40Ca_negative_parity} (negative parity) where
denotations are the same as in Table~\ref{table:208Pb_positive_parity},
but different notation for phonons.
The correspondence of the letters to phonons and the structure
of the RPA phonons are shown in Appendix, Table~\ref{table:40Ca_phonons}.
\begin{table}[h]
\caption{\label{table:40Ca_positive_parity}
The same as in Table \ref{table:208Pb_positive_parity} but
for the positive parity states in $^{40}$Ca.
The letters next to the theoretical values denote phonons:
see Appendix, Table~\ref{table:40Ca_phonons}.
The experimental data are taken from Ref.~\cite{Chen_2017}.}
\begin{ruledtabular}
\begin{tabular}{ccc|ccc}
$I^\pi_n$ & \multicolumn{2}{c|}{$E$[MeV]} & $I^\pi_n$ & \multicolumn{2}{c}{$E$[MeV]} \\[\smallskipamount]
         & Exp.    & SV-bas$_{-0.44}$  &  & Exp. & SV-bas$_{-0.44}$ \\
\hline
\vspace{-2mm}
         &        &          &         &       &       \\
$2^+_2$  & 5.249  & 6.75 aa  & $1^+_1$ &       & 8.25 ac \\
$2^+_4$  & 6.422  & 8.25 ac  & $1^+_1$
                                       &       & 8.50 ad \\
$4^+_2$  & 6.507  & 6.75 aa  & $3^+$   & 6.030 \footnotemark[2]
                                               & 8.25 ac \\
$3^+,4^+$& 7.446  & 8.25 ac  & $3^+$   &       & 8.38 ad \\
$4^+$    & 7.561  & 8.50 ad  & $(5^+)$ & 7.397 \footnotemark[2]
                                               & 8.25 ac  \\
$(6^+)$  & 7.676  & 6.75 aa  & $5^+$   &       & 8.38 ad \\
$(6^+)$  & 8.701  & 8.25 ac  & $(7^+)$ & 8.936 & 8.25 ac \\
$8^+$    & 8.100 \footnotemark[1]
                  & 8.38 ad  & $7^+$   &       & 8.38 ad \\
$8^+$    &        & 9.73 cc  & $(9^+)$ & 11.70 & 9.87 cd  \\
$10^+$   & 11.00 \footnotemark[1]
                  & 13.5 aaaa& $9^+$   &       & 10.12 ce \\
$(10^+)$ & 12.59  & 15.0 aaac& $(11^+)$& 13.53 \footnotemark[2]
                                               & 15.0 aaac \\
$(10^+)$ & 13.19  &15.13 aaad& $11^+$  &       & 15.1 aaad \\
$(12^+)$ & 13.12 \footnotemark[1]
                  & 15.0 aaac& $(13^+)$& 15.15 \footnotemark[1]
                                               & 15.1 aaad \\
$(12^+)$ & 15.75  & 15.1 aaad& $(13^+)$& 16.58 \footnotemark[2]
                                               & 16.5 aacc \\
$(14^+)$ & 17.70  & 16.9 aade& $(15^+)$& 19.19 & 18.13 aadf \\
$(14^+)$ & 18.05  & 17.0 aaee& $(15^+)$&       & 18.25 aaef\\
$(14^+)$ & 18.72  & 18.0 aacf&         &       &       \\
\end{tabular}
\end{ruledtabular}
\footnotetext[1]{$\gamma$ sequence based on $8^+$ \cite{Chen_2017}, p. 211.}
\footnotetext[2]{$3^+$ band \cite{Chen_2017}, p. 211.}
\end{table}
\begin{table}[h]
\caption{\label{table:40Ca_negative_parity}
The same as in Table \ref{table:40Ca_positive_parity} but
for the negative parity states in $^{40}$Ca.}
\begin{ruledtabular}
\begin{tabular}{ccc|ccc}
$I^\pi_n$ & \multicolumn{2}{c|}{$E$[MeV]} & $I^\pi_n$ & \multicolumn{2}{c}{$E$[MeV]} \\[\smallskipamount]
         & Exp.    & SV-bas$_{-0.44}$  &  & Exp. & SV-bas$_{-0.44}$ \\
\hline
         &        &           &         &       &       \\
$(0,1,2)^-$& 8.359\footnotemark[1]
                  & 8.40      & $1^-$   & 5.903 & 7.62 l \\
$2^-$    & 6.025  & 5.33 d    & $3^-$   & 3.737 & 3.38 a \\
$2^-$    & 6.750  & 6.29 f    & $(3^-)$ & 6.160 & 5.25 b \\
$4^-$    & 5.613  & 4.87 c    & $3^-$   & 6.285 & 5.70 i \\
$4^-$    &        & 5.12 e    & $3^-$   & 6.582 & 7.14 j \\
$(6^-)$  & 8.701  & 10.13 aaa &$(2^-,3,4^+)$& 7.623 \footnotemark[2] & 7.45 k \\
$6^-$    &        & 11.62 aac & $5^-$   & 4.491 & 5.00 d \\
$8^-$    & 10.47  & 11.62 aac & $5^-$   & 6.938 \footnotemark[3]& 6.37 f\\
$8^-$    &        & 11.76 aad & $(7^-)$ & 9.033? \footnotemark[4]& 10.13 aaa \\
$(10^-)$ & 13.19  & 11.76 aad & $7^-$   & 11.69 & 11.62 aac\\
$10^-$   &        & 11.87 aae & $(9^-)$ & 10.89 \footnotemark[4]& 11.62 aac \\
$12^-$   &        & 13.25 acd & $(11^-)$& 12.92 \footnotemark[4]& 11.76 aad\\
$12^-$   &        & 13.50 ade & $11^-$  &       & 13.11acc \\
$14^-$   &        & 16.25 cdf & $(13^-)$& 15.31 \footnotemark[4]& 14.8 adf \\
$14^-$   &        &16.38 ddf  & $(13^-)$& 16.58 \footnotemark[5]& 15.0 cde \\
         &        &           & $(15^-)$& 18.21  \footnotemark[4]& 18.51 aaaad \\
\end{tabular}
\end{ruledtabular}
\footnotetext[1]{The experimental assignment to the 8.359 level
is $(0, 1, 2)^-$ \cite{Chen_2017} but our calculations show that the most
probable assignment is $0^-_1$: see the text.}
\footnotetext[2]{The experimental assignment to the 7.623 level
is $(2^-,3,4^+)$ \cite{Chen_2017} but our calculations show that the most
probable assignment is $3^-_5$.}
\footnotetext[3]{The experimental assignment to the 6.938 level
is ($1^-$ to $5^-$) \cite{Chen_2017} but our calculations show that the most
probable assignment is $5^-$.}
\footnotetext[5]{A possible interpretation of the level as a member of the
band as $Kp=0^-$ band and some doubts on this interpretation are given
in Ref.~\cite{Chen_2017}, p. 211.}
\footnotetext[4]{$3^+$ band \cite{Chen_2017}, p. 211.}
\end{table}
The calculations were performed with the same SV-bas$_{-0.44}$
Skyrme parametrization that was used for $^{208}$Pb.
For $^{40}$Ca, the renormalization of phonons changes the phonon energies
by 0.2 -- 0.6 MeV (see Appendix, Table~\ref{table:40Ca_phonons})
that is less than the renormalization for $^{208}$Pb. But many states in $^{40}$Ca
have 3-, 4- and even 5-phonon configurations therefore the renormalization
can change the state energies by 1--1.5 MeV.

When evaluating the quality of the description of levels,
two circumstances should be taken into account.
First of all, our calculations are a first attempt to describe
all these level in a self-consistent model.
As shown in Introduction, there are calculations only for the ND
and SD bands and all the calculations overestimate the energy
of the band-heads by 2 MeV or more, therefore, in many papers,
this energy or an energy of another band state
(e.g., the $I = 16$ state in Ref.~\cite{Wang_2020}) is taken
as a reference.
Secondly, we include in the consideration three rotational
bands that have the following possible interpretation
(see Ref.~\cite{Chen_2017}, p. 211 and references therein):
$\gamma$ sequence based on $8^+$ state, $3^+$ band, and $Kp=0^-$ band
(denotations \cite{Chen_2017}).
The most deviation of our results from the data are just for the states
of these bands but the deviation does not exceed 2 MeV and
does not exceed the above mentioned deviation for the ND
and SD band.
Nevertheless, for some states of the three bands our results are
in general satisfactory
and may be considered as another possible
interpretation of the states.

The one-phonon $E(0^-_1)$ and $E(1^-_1)$ values exceed the proton
separation energy in $^{40}$Ca calculated with the given parameter set,
$S_\mathrm{theor.}(p)=7.565$ MeV
(the experimental value $S_\mathrm{exp.}(p)=8328.17$ \cite{Chen_2017}).
These energies have been recalculated in RenTBA taking
into account the s.p. continuum, and the values given in the tables
takes this effect into account.
The continuum effect is not significant for the two- and many-phonon
states.

The calculation results allow us to refine the identification
of some levels in $^{40}$Ca.
The experimental assignment to the 7.623 level is
$(2^-,3,4^+)$ \cite{Chen_2017}. Possible theoretical states
corresponding to this level are $2^-_3$, $3^-_5$,  $3^+_2$, and $4^+_7$
with the energies 8.77, 7.45 k, 8.38 ad and 8.63 ab, respectively.
It should be noted that the $4^+_1$ and $4^+_3$ states belong
to the ND and SD bands, respectively, and therefore they
are not shown in the table. The most probable assignment
for the 7.623 level is $3^-_5$.

The experimental assignment to the 6.938 level
is ($1^-$ to $5^-$) \cite{Chen_2017} but our calculations show that the most
probable assignment is $5^-$ because all the other states have too high energies.
The experimental assignment to the 8.359 level is $(0, 1, 2)^-$ \cite{Chen_2017}.
Our calculations show that the most probable assignment is $0^-_1$.

%------------------------------
\subsection{Semi-magic $^{90}$Zr}%\nonumber
%------------------------------

New experimental data for the states up to spin $I=20$ in $^{90}$Zr
and the shell-model calculations are given in Ref.~\cite{Dey_2022}.
The $^{90}$Zr high-spin states were interpreted to be generated
by the recoupling of stretched proton and neutron configurations.
The shell-model calculations ~\cite{Dey_2022} used the fitted s.p. energies
and two-body matrix elements of the effective interaction.
Besides the two bands considered in Ref.~\cite{Dey_2022}, there are many
other known levels in $^{90}$Zr having both the low and high spins.
The calculations presented in this subsection are the first attempt
to describe all known states of $^{90}$Zr within the framework
of a self-consistent method.

The results for $^{90}$Zr are shown in Tables~\ref{table:90Zr_positive_parity}
(positive parity states) and \ref{table:90Zr_negative_parity}
(negative parity) where denotations are the same as
in Table~\ref{table:208Pb_positive_parity}, but different notation for phonons.
The correspondence of the letters to phonons and the structure
of the RPA phonons are shown in Appendix, Table~\ref{table:90Zr_phonons}.
The experimental data are taken from Refs.~\cite{NDS_2020_A=90,Dey_2022}.

It should be noted that there are many unassigned levels above 3.5 MeV,
such as 3.557 MeV, 3.9324 MeV, etc. \cite{NDS_2020_A=90}, so it is possible
that some of the lowest assigned levels are not yrast or yrare.
This should be taken into account when comparing theory with data,
since the theoretical values are given just for the yrast
and yrare states.

\begin{table}[h]
\caption{\label{table:90Zr_positive_parity}
The same as in Table \ref{table:208Pb_positive_parity} but
for the positive parity states in $^{90}$Zr.
The letters next to the theoretical values denote phonons:
see Appendix, Table~\ref{table:90Zr_phonons}.
The experimental data are taken from Ref.~\cite{NDS_2020_A=90,Dey_2022}.
The theoretical results were obtained with SV-bas$_{-0.44}$
and SLy4 Skyrme EDF parametrizations but, for SV-bas$_{-0.44}$ (denoted as bas$_{-0.44}$),
the energy of the single-particle $\pi 1g_{9/2}$ state
was increased by 0.6 MeV in these calculations.
}
\begin{ruledtabular}
\begin{tabular}{cccc|cccc}
$I^\pi_n$& \multicolumn{3}{c|}{$E$[MeV]} & $I^\pi_n$ & \multicolumn{3}{c}{$E$[MeV]} \\
[\smallskipamount]
        & exp.   & bas$_{-0.44}$
                         & SLy4    &         & exp. & bas$_{-0.44}$
                                                                & SLy4   \\
\hline
\vspace{-2mm}
        &       &       &          &          &      &       &         \\
$0^+$   & 1.76  & 3.67 \footnotemark[24]
                        & 4.27 aa  \footnotemark[24]
                                   & $1^+$    &      & 3.67  & 4.28 aa \\
$0^+$   & 4.12  & 5.03  & 5.68 ab  & $1^+$    &      & 5.96  & 5.75 ab \\
$0^+$   & 4.43  & 5.19  & 5.77 bb  &          &      &       &  \\
$2^+$   & 2.19  & 3.60 \footnotemark[24]
                        & 4.27 aa \footnotemark[24]
                                   & $(3^+)$  & 4.26 & 3.67  & 4.28 aa \\
$2^+$   & 3.31  & 3.67  & 4.31 d   & $3^+$    &      & 3.84  & 4.94 d \\
$2^+$   & 3.84  & 4.93  & 5.46 ab  &          &      &       &  \\
$2^+$   & 4.22  & 5.81  & 5.77 bb  &          &      &       &  \\
$2^+$   & 4.23  & 5.90  & 5.97 ac  &          &      &       &  \\
$4^+$   & 3.08  & 3.67  & 4.27 aa  & $(5^+)$  & 4.45 & 3.67  & 4.28 aa \\
$4^+$   & 4.06  & 3.68  & 4.70 d   & $5^+$    & 4.87 & 3.81  & 4.89 d \\
$4^+$   & 4.30  & 5.02  & 5.02 ab  & $7^+$    & \footnotemark[1]
                                                     & 3.67  &  4.28 aa \\
$4^+$   & 4.33  & 5.25  & 5.77 bb  & $7^+$    & 5.06 & 4.03  & 4.75d \\
$(4^+)$ & 4.35  & 5.89  & 5.79 ab  & $9^+$    & 5.25 & 3.67  & 4.28 aa \\
$4^+$   & 4.47  & 5.90  & 6.18 ac  & $9^+$    & 5.79 & 5.08  & 5.83 ab  \\
$6^+$   & 3.45  & 3.62  & 4.27 aa  & $11^+$   & 6.28 & 6.10  & 6.12 ac \\
$6^+$   &       & 3.74  & 4.88 d   & $(11^+)$ & 7.19 & 6.53  & 7.42 bb \\
$8^+$   & 3.59  & 3.67  & 4.27 aa  & $13^+$   & 7.44 & 7.46  & 8.01 cc \\
$8^+$   & 5.16  & 5.00  & 5.02 ab  & $13^+$   &      & 8.36  & 9.10 aad  \\
$10^+$  & 5.64  & 5.11  & 5.83 ab  & $15^+$   & 8.95 & 8.56  & 9.24 aad \\
$10^+$  & 7.03  & 5.84  & 6.10 ac  & $15^+$   & 9.33 & 8.93  & 10.4 abd \\
$(12+)$ & 6.77  & 7.44  & 7.77 bc  & $(15)^+$ & 9.84 & 9.54  & 10.9 acd\\
$12^+$  & 7.22  & 7.49  & 7.99 cc  & $(17^+)$ & 10.8 & 9.10  & 10.6 abd\\
$14^+$  & 8.06  & 7.52  & 9.07 aad & $17^+$   &      & 9.92  & 10.9 acd\\
$14^+$  &       & 8.85  &10.2 abd &$(19^+)$   & 12.1 & 11.6  & 12.5 bcd  \\
$16^+$  & \footnotemark[2]
                & 7.65  & 8.90 aad & $19^+$   &      & 11.5  & 12.6 aabd \\
$16^+$  & 10.1  & 8.92  & 10.6 abd &          &      &       & \\
$(18^+)$& 11.4  & 10.1  & 10.9 acd &          &      &       & \\
$18^+$  &       & 10.7  & 12.2 bbc &          &      &       & \\
$(20^+)$& 13.0  & 11.5 \footnotemark[3]
                        & 12.8 ccd &          &      &       & \\
$20^+$  &        & 12.7 \footnotemark[3]
                         &13.7 aadd&          &       &       & \\
\end{tabular}
\end{ruledtabular}
\footnotetext[1]{The $7^+$ state belongs to the 'aa' multiplet
where the known $2^+_1$, $4^+_1$, $6^+_1$, and $8^+_1$ members
have energies of 2.5 -- 3.5 MeV so there should be a $7^+$ state
with an energy of 3 -- 4 MeV. The same is valid for the 'd' multiplet.}
\footnotetext[2]{The level 10.12584 \cite{NDS_2020_A=90} cannot be the first state
of $16^+$; it's either 2nd or 3rd: see the text.}
\footnotetext[3]{For SV-bas$_{-0.44}$, the first and second have the
configurations 'aadd' and 'ccd', respectively.}
\footnotetext[24]{The first $0^+$ and $2^+$ state are probably deformed and therefore
outside the scope of the model: see the text.}
\end{table}
\begin{table}[h]
\caption{\label{table:90Zr_negative_parity}
The same as in Table \ref{table:90Zr_positive_parity} but
for the negative parity states in $^{90}$Zr.
}
\begin{ruledtabular}
\begin{tabular}{cccc|cccc}
$I^\pi_n$& \multicolumn{3}{c|}{$E$[MeV]} & $I^\pi_n$ & \multicolumn{3}{c}{$E$[MeV]} \\
[\smallskipamount]
        & exp. & bas$_{-0.44}$
                      & SLy4    &         & exp. & bas$_{-0.44}$
                                                         & SLy4   \\
\hline
\vspace{-2mm}
        &      &      &          &         &      &      &         \\
$0^-$   &      & 5.59 & 6.93 ad  & $1^-$   &      & 5.60 & 5.00 ad \\
$0^-$   &      & 6.75 & 7.79 aab & $1^-$   &      & 6.79 & 5.59 bd \\
$2^-$   &      & 4.53 & 4.83 c   & $3^-$   & 2.75 & 2.60 & 2.88 b  \\
$2^-$   &      & 5.60 & 6.88 ad  & $(3^-)$ & 4.50 & 3.84 & 3.97 c  \\
$(4^-)$ & 2.74 & 1.89 & 2.26 a   & $(3^-)$ & 4.81 & 5.62 & 5.86 h  \\
$(4^-)$ & 4.22 & 3.24 & 3.78 b   & $3^-$   & 5.63 & 6.01 & 6.66 ad \\
$(4^-)$ & 4.54 & 4.00 & 4.10 c   & $3^-$   & 5.67 & 6.66 & 7.46 aab \\
$(4^-)$ & 4.94 & 5.75 & 6.82 ad  & $5^-$   & 2.32 & 1.78 & 2.02 a  \\
$6^-$   & 4.23 & 3.38 & 3.76 b   & $5^-$   & 3.96 & 3.14 & 3.66 b  \\
$6^-$   &      & 3.87 & 3.95 c   & $(5^-)$ & 4.30 & 3.90 & 3.88 c \\
$8^-$   &      & 5.64 & 6.94 ad  & $7^-$   & 4.37 & 4.49 & 4.06 c \\
$8^-$   &      & 6.72 & 7.64 bd  & $7^-$   &      & 5.63 & 6.84 ad \\
$10^-$  & 6.38 & 5.70 & 6.96 ad  & $9^-$   &      & 5.66 & 6.97 ad \\
$10^-$  & 6.72 & 6.72 & 7.64 bd
 \footnotemark[1]                & $9^-$   &      & 6.66 & 7.55 aab \\
$(12-)$ &      & 5.81 & 6.77 ad  & $11^-$  & 6.95 & 5.75 & 6.89 ad  \\
$12^-$  &      & 6.90 & 7.79 aab & $11^-$  & 7.01 & 6.77 & 7.53 aab \\
$14^-$  & 8.86 & 7.78 & 8.22 aac & $13^-$  &      & 6.91 & 7.81 aab \\
$14^-$  &      & 8.40 & 8.81 cd  & $13^-$  &      & 7.41 & 8.50 bd \\
$(16^-)$& 9.71 & 10.1 & 10.1 acc & $15^-$  & 8.96 & 9.10 & 9.63 abc \\
$(16^-)$& 10.0 & 10.6 & 11.5 bbc & $15^-$  & 9.83 & 9.87 & 10.10 acc \\
$16^-$  & 10.1 & 10.7 & 11.6 bcc & $15^-$  & 9.97 & 10.3 & 10.63 bcc \\
$16^-$  & 10.4 & 11.1 & 11.7 add & $17^-$  & 10.8 & 9.58 & 11.69 add \\
$18^-$  & 11.3 & 9.55 & 11.6 add & $17^-$  & 10.9 & 10.6 & 11.69 bcc \\
$18^-$  & 11.4 & 11.2 & 12.0 ccc &         &      &      &  \\
$20^-$  & 12.6 & 12.2 & 12.8 aacd& $19^-$  & 12.1 & 11.2 & 12.5 aabd \\
$20^-$  & 13.0 & 12.3 & 13.7 cdd & $19^-$  & 12.3 & 12.1 & 12.8 aacd \\
\end{tabular}
\end{ruledtabular}
\footnotetext[1]{The 'bd' and 'aab' configurations have very close energies
for the $8^-_2$ level; the same is also for $10^-_2$.}
\end{table}

The calculations with the SV-bas$_{-0.44}$ parameter set underestimate
energies of many states in $^{40}$Ca by 2 -- 4 MeV, particularly
for the high-spin states. At the same time, it turns out that a small
change in only one s.p. energy, for the $\pi 1g_{9/2}$ state, significantly
improves agreement with experiment.
The SV-bas$_{-0.44}$ results given in Tables~\ref{table:90Zr_positive_parity}
and \ref{table:90Zr_negative_parity} were obtained with the energy of
the $\pi 1g_{9/2}$ state increased by 0.6 MeV. This feature of the calculations
is designated in Tables~\ref{table:90Zr_positive_parity} and \ref{table:90Zr_negative_parity}
as bas$_{-0.44}$.

To keep full self-consistency of the method, we also performed
calculations for $^{90}$Zr with the well known forces SLy4.
It is interesting to note, that the self-consistent SLy4 results
for many levels are rather close to the bas$_{-0.44}$ values calculated
with this increased $\pi 1g_{9/2}$ energy.

The result for $^{90}$Zr presented in the tables are
in general acceptable,
particularly, taking into account the high sensitivity to the EDF
parameters.
The most significant deviations are for the first excited $0^+$
and $2^+$ states, i. e. $0^+_2$ and $2^+_1$, but these states have,
probably, a large deformation.
The experimental ratio
\be
\frac{E(4^+_1)-E(0^+_2)}{E(2^+_1)-E(0^+_2)} = 3.07
\ee
is very close to the rotational limit 3.33.
The deformed states are beyond the scope of the model using
the spherical basis but
for completeness we include the $0^+$ levels of $^{90}$Zr in the table.

At the same time, unlike $^{40}$Ca, the deformation manifested in the
$0^+_2$, $2^+_1$ and, may be, $4^+_1$ states in $^{90}$Zr is not seen
in the $6^+_1$ state, since the experimental ratio
\be
\frac{E(6 ^+_1 )-E(0^+_2)}{E(2^+_1)-E(0^+_2)} = 3.93
\ee
is significantly less than the rotational value of 7.00.
All the result for $^{90}$Zr, excluding $0^+_2$ and $2^+_1$ states,
are in a better and sometimes in nice agreement with the data.

One can expect that these results can serve as a guideline for the search of the
new states in $^{90}$Zr.
For example, the $6^+_1$, $7^+_1$, and $8^+_1$ levels
are a part of the 'aa' multiplet where the  $6^+_1$ and $8^+_1$ states
have the experimental values of energies are 3.45--3.59 MeV,
so there should be a state $7^+$, which has approximately the same energy
and this state should be yrast. Therefore the experimental 5.060 level is
placed as the second $7^+$ state in Table~\ref{table:90Zr_positive_parity}.
The states $14^+$ and $16^+$ are a part of the "aad" multiplet,
so there must be a level $16^+_1$ that is close in energy to the level
$14^+$ with $E(14^+) = 8.058$ MeV.
Therefore, 10.12584 cannot be the yrast state of $16^+$;
it's either 2nd or 3rd state.

%\clearpage

%---------
\section{SUMMARY AND CONCLUSIONS}
%---------

In the paper, the self-consistent calculations of the structure and energies
of the high-spin states in doubly magic $^{208}$Pb and $^{40}$Ca and
in semi-magic $^{90}$Zr have been presented.
The calculations were performed within the framework of the multiphonon model
including the renormalized phonons in the harmonic approximation
and based on the energy-density functional (EDF) of the Skyrme type.
The basic elements of the theory, the phonons, are determined
within the renormalized time-blocking approximation which is a non-linear
version of the model of the beyond-RPA type.

In $^{208}$Pb, the calculations have been performed for all
the experimentally known levels of spins
$I \ge 13$  and, when there are no data,
for all yrast and yrare states of $13 \le I \le 30$.
In $^{40}$Ca and $^{90}$Zr, the results have been obtained for all the
experimentally known high-spin states which can be described within our approach
based on the spherically-symmetric mean field.
The results for $^{208}$Pb are in fairly well agreement with the experimental data
despite the fact that no new fitting parameters in the underlying Skyrme EDF
were introduced.
The results for $^{40}$Ca and $^{90}$Zr are less satisfactory with respect to
comparison with data, but in part it is explained by the parametrization
of the Skyrme EDF used in all the calculations and adjusted previously
to the properties of the excited states of $^{208}$Pb.

The choice of the EDF is important in describing the multiphonon states
because of their sensitivity to the parameters of the EDF which affect
the phonon's energies.
This is demonstrated in the case of $^{90}$Zr.
A change in one phonon causes
much more significant change in the n-phonon energy.
At the same time, there are many high-spin states having
rather simple structure in terms of the renormalized
phonons. Therefore the energies of the multiphonon states can be used
as the additional fit data for adjustment of the parameters of the EDFs.

Our results confirm the conclusion of Ref. \cite{Lyutorovich_2022}
about importance of the use of the renormalized phonons in the description
of the high-spin states. The renormalization reduces the phonon's energies
that in the most cases improves agreement with data.

\begin{acknowledgements}
This work was supported by the Russian Foundation for Basic Research,
project number 21-52-12035.  This research was carried out using
computational resources provided by the Computer Center of
St. Petersburg State University.
\end{acknowledgements}

%---------
\appendix
%---------

%---------
\section{Phonons}
%---------

Structure and energies of the phonons used in the calculations of the multiphonon
states in $^{208}$Pb are shown in Table~\ref{table:208Pb_phonons}
for the SV-bas$_{-0.44}$ Skyrme parameter set.
Here, the last two columns show the energies of the RPA and RenTBA (renormalized) phonons.
The letters a, b, c, ... denote multiplets of phonons having
approximately the same structure.
%The renormalization adds to the phonons many complex excitations
%of the type 2p2h, 3p3h, etc., but the impurity amplitudes are not
%very large and are therefore not shown here.

Tables~\ref{table:40Ca_phonons} and \ref{table:90Zr_phonons}
show the structure and energies of the phonons for $^{40}$Ca and $^{90}$Zr,
respectively.
\begin{table*}
\caption{\label{table:208Pb_phonons} Structure and energy (in MeV)
of some phonons appearing in the n-phonon states of $^{208}$Pb.
The last three columns show the experimental data \cite{NDS_2007,Heusler_2016}
and the energies of the RPA and RenTBA (renormalized) phonons.
The letters a, b, c, ... denote multiplets of phonons having
approximately the same structure.
}
\begin{ruledtabular}
\begin{tabular}{rclccc}
   & $I^\pi$    & Configuration                                      & \multicolumn{3}{c}{$E$[MeV]} \\
                                                                         \cline{4-6}
   &            &                                                    &      & \vspace{-2mm}  &  \\
   &            &                                                    & Exp. & RPA & RenTBA \\
\hline
\vspace{-2mm}
\phantom{aaaaa}&&                                                    &   &      &      \\
 a & $3^-_{  1}$ & $\nu\, 2g_{ 9/2} \, 3p_{ 3/2}{}^{-1}\,21\%
                + \pi\, 1h_{ 9/2} \, 2d_{ 3/2}{}^{-1}\,20\% + \dots$ & 2.614 & 3.10 & 2.87  \\[\medskipamount]
 b & $2^+_{  1}$ & $\nu\, 2g_{ 9/2} \, 1i_{13/2}{}^{-1} \,63\%
               + \pi\, 2f_{ 7/2} \, 1h_{11/2}{}^{-1} \,19\% $        & 4.086 & 4.42 & 4.01 \\
   & $4^+_{  1}$ & $\nu\, 2g_{ 9/2} \, 1i_{13/2}{}^{-1}\, 58\%
                + \pi\, 1h_{ 9/2} \, 1h_{11/2}{}^{-1}\, 16\%$        & 4.324 & 4.80 & 4.31  \\
   & $6^+_{  1}$ & $\nu\, 2g_{ 9/2} \, 1i_{13/2}{}^{-1}\,64\%
                 + \pi\, 1h_{ 9/2} \, 1h_{11/2}{}^{-1}\,18\%$        & 4.424 & 5.13 & 4.56 \\[\medskipamount]
 c & $5^+_{  1}$ & $\nu\, 2g_{ 9/2} \, 1i_{13/2}{}^{-1}\,99\% $      & 4.962 & 5.36 & 4.69  \\
   & $7^+_{  1}$ & $\nu\, 2g_{ 9/2} \, 1i_{13/2}{}^{-1}\,98\% $      & 4.867 & 5.36 & 4.68  \\
   & $8^+_{  1}$ & $\nu\, 2g_{ 9/2} \, 1i_{13/2}{}^{-1} \,92\%$      & 4.611 & 5.31 & 4.66  \\
   & $9^+_{  1}$ & $\nu\, 2g_{ 9/2} \, 1i_{13/2}{}^{-1} \,99\%$      & 5.010 & 5.39 & 4.70  \\
   & $10^+_{  1}$ & $\nu\, 2g_{ 9/2} \, 1i_{13/2}{}^{-1}\, 98\% $    & 4.895 & 5.34 & 4.65  \\
   & $11^+_{  1}$ & $\nu\, 2g_{ 9/2} \, 1i_{13/2}{}^{-1} \, 100\%$   & 5.235 & 5.54 & 4.85  \\[\medskipamount]
 d & $4^-_{  1}$ & $\nu\, 2g_{ 9/2} \, 3p_{ 1/2}{}^{-1} \, 98\%$     & 3.475 & 4.15 & 3.65  \\
   & $5^-_{  1}$ & $\nu\, 2g_{ 9/2} \, 3p_{ 1/2}{}^{-1}\,70\%
            + \pi\, 1h_{ 9/2} \, 3s_{ 1/2}{}^{-1}\,18\%$             & 3.198 & 3.92 & 3.51  \\[\medskipamount]
 e & $4^-_{  2}$ & $\pi\, 1h_{ 9/2} \, 3s_{ 1/2}{}^{-1}\,99\%$       & 3.947 & 4.36 & 3.93 \\
   & $5^-_{  2}$ & $\pi\, 1h_{ 9/2} \, 3s_{ 1/2}{}^{-1}\,70\%
            + \nu\, 2g_{ 9/2} \, 3p_{ 1/2}{}^{-1}\,24\%$             & 3.708 & 4.29 & 3.85 \\[\medskipamount]
 f & $5^-_{  3}$ & $\nu\, 2g_{ 9/2} \, 3p_{ 3/2}{}^{-1}\,48\%
            + \pi\, 1h_{ 9/2} \, 2d_{ 3/2}{}^{-1}\,28\%$             & 3.961 & 4.73 & 4.21 \\[\medskipamount]
 g & $6^-_{  2}$ & $\pi\, 1h_{ 9/2} \, 2d_{ 3/2}{}^{-1}\,83\%$       & 4.206 & 5.05 & 4.58 \\[\medskipamount]
 h & $6^-_{  3}$ & $\nu\, 2g_{ 9/2} \, 2f_{ 5/2}{}^{-1}\,79\%$       & 4.383 & 5.14 & 4.52 \\
   & $7^-_{  1}$ & $\nu\, 2g_{ 9/2} \, 2f_{ 5/2}{}^{-1}\,97\%$       & 4.037 & 5.41 & 4.75 \\[\medskipamount]
 i & $7^-_{  2}$ & $\nu\, 1i_{11/2} \, 3p_{ 3/2}{}^{-1}\,96\%$       & 4.680 & 6.00 & 5.44 \\[\medskipamount]
 j & $8^-_{  1}$ & $\nu\, 1i_{11/2} \, 2f_{ 5/2}{}^{-1}\,99\%$       & 4.919 & 6.29 & 5.70 \\[\medskipamount]
 k & $8^+_{  2}$ & $\pi\, 1h_{ 9/2} \, 1h_{11/2}{}^{-1}\,76\%$       & 4.861 & 5.74 & 5.15 \\
   & $9^+_{  2}$ & $\pi\, 1h_{ 9/2} \, 1h_{11/2}{}^{-1}\,99\%$       & 5.162 & 5.73 & 5.15 \\
   & $10^+_{  2}$ & $\pi\, 1h_{ 9/2} \, 1h_{11/2}{}^{-1}\,79\%$      & 5.069 & 6.11 & 5.53 \\[\medskipamount]
 l & $9^+_{  3}$ & $\nu\, 1i_{11/2} \, 1i_{13/2}{}^{-1} \,98\%$      & 5.327 & 6.48 & 5.82 \\
   & $10^+_{  3}$ & $\nu\, 1i_{11/2} \, 1i_{13/2}{}^{-1}\,82\%
            + \pi\, 1h_{ 9/2} \, 1h_{11/2}{}^{-1} \,18\%$            & 5.537 & 6.72 & 6.01 \\
   & $11^+_{  2}$ & $\nu\, 1i_{11/2} \, 1i_{13/2}{}^{-1} \,100\%$    & 5.750 & 6.46 & 5.80 \\
   & $12^+_{  1}$ & $ \nu\, 1i_{11/2} \, 1i_{13/2}{}^{-1} \,100\%$   & 5.864 & 7.05 & 6.37 \\[\medskipamount]
 m & $9^+_{  4}$ & $\pi\, 2f_{ 7/2} \, 1h_{11/2}{}^{-1} \,94\%$      & 5.901 & 6.69 & 5.79 \\[\medskipamount]
 o & $7^-_{  6}$ & $ \pi\, 1i_{13/2} \, 1h_{11/2}{}^{-1} \,75\%
            + \nu\, 1j_{15/2} \, 1i_{13/2}{}^{-1}\,21\%$\,, \
      $X_1 > 0, X_2 > 0$                                             &       & 7.48 & 6.51 \\
   & $9^-_{  1}$ & $\pi\, 1i_{13/2} \, 1h_{11/2}{}^{-1} \,99\%$      & 6.861 & 7.51 & 6.56 \\
   & $11^-_{  1}$ & $\pi\, 1i_{13/2} \, 1h_{11/2}{}^{-1} \,96\%$     &       & 7.48 & 6.52 \\[\medskipamount]
 p & $8^-_{  2}$ & $\pi\, 1i_{13/2} \, 1h_{11/2}{}^{-1} \,59\%
            + \nu\, 1j_{15/2} \, 1i_{13/2}{}^{-1}\,37\%$\,, \
     $X_1 > 0, X_2 > 0$                                              & 5.836 & 7.33 & 6.40 \\
   & $10^-_{  1}$ & $\pi\, 1i_{13/2} \, 1h_{11/2}{}^{-1} \,53\%
            + \nu\, 1j_{15/2} \, 1i_{13/2}{}^{-1}\,47\%$\,, \
     $X_1 > 0, X_2 > 0$                                              & 6.283 & 7.39 & 6.42 \\[\medskipamount]
 q & $12^-_{  1}$ & $\nu\, 1j_{15/2} \, 1i_{13/2}{}^{-1}\, 68\%
             + \pi\, 1i_{13/2} \, 1h_{11/2}{}^{-1}\, 32\%$           & 6.435 & 7.52 & 6.49 \\[\medskipamount]
 r & $11^-_{  2}$ & $\nu\, 1j_{15/2} \, 1i_{13/2}{}^{-1} \,96\%$     &       & 7.60 & 6.50 \\
   & $13^-_{  1}$ & $\nu\, 1j_{15/2} \, 1i_{13/2}{}^{-1}\, 100\%$    &       & 7.57 & 6.47 \\
   & $14^-_{  1}$ & $\nu\, 1j_{15/2} \, 1i_{13/2}{}^{-1}\, 100\%$    & 6.743 & 8.02 & 6.81 \\[\medskipamount]
 s & $12^-_{  2}$ & $\pi\, 1i_{13/2} \, 1h_{11/2}{}^{-1} \,68\%
            + \nu\, 1j_{15/2} \, 1i_{13/2}{}^{-1}\,32\%$\,, \
     $X_1 > 0, X_2 < 0$                                              & 7.061 & 8.06 & 6.92 \\
t & $6^-_{  4}$ & $\nu\, 1i_{11/2} \, 3p_{ 1/2}{}^{-1} \,82\%$       & 4.481 & 5.21 & 4.72 \\
u & $8^+_{  3}$ & $\nu\, 1j_{15/2} \, 3p_{ 1/2}{}^{-1} \,83\%$       & 5.093 & 6.19 & 5.32 \\
v & $12^+_{  2}$ & $\nu\, 1j_{15/2} \, 1h_{ 9/2}{}^{-1} \,83\%$      & \footnotemark[1]
                                                                             & 10.37 & 8.95 \\
\end{tabular}
\end{ruledtabular}
\footnotetext[1]{The $12^+_{  2}$ state has a 2-phonon confuguration and therefore
its energy should not be compared with the 1-phonon energy.}
\end{table*}
\begin{table*}
\caption{\label{table:40Ca_phonons}
The same as in Table \ref{table:208Pb_phonons} but for $^{40}$Ca.
The experimental data are taken from Ref.~\cite{Chen_2017}.
}
\begin{ruledtabular}
\begin{tabular}{rclccc}
   & $I^\pi$     & Configuration                                      & \multicolumn{3}{c}{$E$[MeV]} \\
                                                                         \cline{4-6}
   &             &                                                    &       & \vspace{-2mm}  &  \\
   &             &                                                    & Exp.  & RPA & RenTBA \\
\hline
\vspace{-2mm}
\phantom{aaaaa}& &                                                    &       &      &      \\
 a & $3^-_{  1}$ & $\pi\, 1f_{ 7/2} \, 1d_{ 3/2}{}^{-1}\, 37\%
                + \nu\, 1f_{ 7/2} \, 1d_{ 3/2}{}^{-1}\, 30\% + \dots $\,, \
     $X_1 > 0, X_2 > 0$                                               & 3.737 & 3.58 & 3.38 \\[\medskipamount]
 b & $3^-_{  2}$ & $\pi\, 1f_{ 7/2} \, 1d_{ 3/2}{}^{-1}\, 52\%
               + \nu\, 1f_{ 7/2} \, 1d_{ 3/2}{}^{-1}\, 29\% $\,, \
     $X_1 > 0, X_2 < 0$                                               & 6.160 & 5.58 & 5.25 \\[\medskipamount]
 c & $4^-_{  1}$ & $\pi\, 1f_{ 7/2} \, 1d_{ 3/2}{}^{-1}\,97\% $       & 5.613 & 5.17 & 4.87 \\[\medskipamount]
 d & $5^-_{  1}$ & $\pi\, 1f_{ 7/2} \, 1d_{ 3/2}{}^{-1}\,60\%
                + \nu\, 1f_{ 7/2} \, 1d_{ 3/2}{}^{-1}\,39\%$\,, \
      $X_1 > 0, X_2 > 0$                                              & 4.491 & 5.33 & 5.00 \\
   & $2^-_{  1}$ & $\pi\, 1f_{ 7/2} \, 1d_{ 3/2}{}^{-1} \,64\%
                + \nu\, 1f_{ 7/2} \, 1d_{ 3/2}{}^{-1} \,36\%$\,, \
     $X_1 > 0, X_2 > 0$                                               & 6.025 & 5.67 & 5.33 \\[\medskipamount]
 e & $4^-_{  2}$ & $\nu\, 1f_{ 7/2} \, 1d_{ 3/2}{}^{-1} \, 96\%$      &       & 5.45 & 5.12 \\[\medskipamount]
 f & $2^-_{  2}$ & $\nu\, 1f_{ 7/2} \, 1d_{ 3/2}{}^{-1}\,61\%
                + \pi\, 1f_{ 7/2} \, 1d_{ 3/2}{}^{-1}\,34\%$\,, \
     $X_1 > 0, X_2 < 0$                                               & 6.750 & 6.71 & 6.29 \\
   & $5^-_{  2}$ & $\nu\, 1f_{ 7/2} \, 1d_{ 3/2}{}^{-1}\,59\%
                + \pi\, 1f_{ 7/2} \, 1d_{ 3/2}{}^{-1}\,38\%$\,, \
     $X_1 > 0, X_2 < 0$                                               & 6.938 & 6.81 & 6.37 \\[\medskipamount]
 i & $3^-_{  3}$ & $\nu\, 1f_{ 7/2} \, 1d_{ 3/2}{}^{-1}\,35\%
                + \nu\, 1f_{ 7/2} \, 2s_{ 1/2}{}^{-1}\,33\% + \dots $\,, \
     $X_1 > 0, X_2 > 0$                                               & 6.285 & 6.02 & 5.69 \\[\medskipamount]
 j & $3^-_{  4}$ & $\nu\, 1f_{ 7/2} \, 2s_{ 1/2}{}^{-1}\,44\%
                + \pi\, 1f_{ 7/2} \, 2s_{ 1/2}{}^{-1}\,35\% + \dots $\,, \
     $X_1 > 0, X_2 < 0$                                               & 6.582 & 7.55 & 7.14 \\[\medskipamount]
 k & $3^-_{  5}$ & $\pi\, 2p_{ 3/2} \, 1d_{ 3/2}{}^{-1}\,57\%
                + \nu\, 2p_{ 3/2} \, 1d_{ 3/2}{}^{-1}\,36\% $\,, \
     $X_1 > 0, X_2 > 0$                                               & 7.623 & 8.08 & 7.45 \\[\medskipamount]
 l & $1^-_{  1}$ & $\pi\, 2p_{ 3/2} \, 1d_{ 3/2}{}^{-1}\,55\%
                + \nu\, 2p_{ 3/2} \, 1d_{ 3/2}{}^{-1}\,25\% $\,, \
     $X_1 > 0, X_2 > 0$                                               & 5.903 & 8.25 & 7.63 \\
\end{tabular}
\end{ruledtabular}
\end{table*}
\begin{table*}
\caption{\label{table:90Zr_phonons} The same as in Table~\ref{table:208Pb_phonons}
but for $^{90}$Zr. The calculation values are given for the SLy4 parameter set.
The experimental data are taken from Refs.~\cite{NDS_2020_A=90,Dey_2022}.
}
\begin{ruledtabular}
\begin{tabular}{rclcccclccc}
   & $I^\pi$     & Configuration                                  & \multicolumn{3}{c}{$E$[MeV]}
   & $I^\pi$     & Configuration                                  & \multicolumn{3}{c}{$E$[MeV]} \\
                                                                     \cline{4-6} \cline{9-11}
   &             & \vspace{-2mm}                                  &   &      &     &    &    &   &   \\
   &             &                                                & Exp. & RPA & RenTBA
   &             &                                                & Exp. & RPA & RenTBA                  \\
\hline
\vspace{-2mm}
\phantom{aaa}&   &                                                &      &      &     &    &    &   &   \\
 a & $4^-_{  1}$ & $\pi\, 1g_{ 9/2} \, 2p_{ 1/2}{}^{-1}\, 100\%$  & 2.74 & 2.49 & 2.26
   & $5^-_{  1}$ & $\pi\, 1g_{ 9/2} \, 2p_{ 1/2}{}^{-1}\, 100\%$  & 2.32 & 2.20 & 2.02 \\[\medskipamount]
 b & $3^-_{  1}$ & $\pi\, 1g_{ 9/2} \, 2p_{ 3/2}{}^{-1}\, 88\%$   & 2.75 & 3.09 & 2.88
   & $4^-_{  2}$ & $\pi\, 1g_{ 9/2} \, 2p_{ 3/2}{}^{-1}\, 98\%$   & 4.22 & 4.22 & 3.78 \\
   & $5^-_{  2}$ & $\pi\, 1g_{ 9/2} \, 2p_{ 3/2}{}^{-1}\, 99\%$   & 3.96 & 4.10 & 3.66
   & $6^-_{  1}$ & $\pi\, 1g_{ 9/2} \, 2p_{ 3/2}{}^{-1}\, 98\%$   & 4.23 & 4.18 & 3.76 \\[\medskipamount]
 c & $2^-_{  1}$ & $\pi\, 1g_{ 9/2} \, 1f_{ 5/2}{}^{-1} \, 99\%$  &      & 5.36 & 4.84
   & $3^-_{  2}$ & $\pi\, 1g_{ 9/2} \, 1f_{ 5/2}{}^{-1} \, 91\%$  & 4.81 & 4.40 & 3.96 \\
   & $4^-_{  3}$ & $\pi\, 1g_{ 9/2} \, 1f_{ 5/2}{}^{-1} \, 98\%$  & 4.54 & 4.56 & 4.10
   & $5^-_{  3}$ & $\pi\, 1g_{ 9/2} \, 1f_{ 5/2}{}^{-1} \, 99\%$  & 4.30 & 4.34 & 3.87 \\
   & $6^-_{  2}$ & $\pi\, 1g_{ 9/2} \, 1f_{ 5/2}{}^{-1} \, 97\%$  &      & 4.39 & 3.95
   & $7^-_{  1}$ & $\pi\, 1g_{ 9/2} \, 1f_{ 5/2}{}^{-1} \, 99\%$  & 4.37 & 4.41 & 4.06 \\[\medskipamount]
 d & $2^+_{  1}$ & $\nu\, 2d_{ 5/2} \, 1g_{ 9/2}{}^{-1} \, 97\%$  &      & 4.67 & 4.31
   & $3^+_{  1}$ & $\nu\, 2d_{ 5/2} \, 1g_{ 9/2}{}^{-1} \, 99\%$  &      & 5.46 & 4.95 \\
   & $4^+_{  1}$ & $\nu\, 2d_{ 5/2} \, 1g_{ 9/2}{}^{-1} \, 99\%$  &      & 5.16 & 4.70
   & $5^+_{  1}$ & $\nu\, 2d_{ 5/2} \, 1g_{ 9/2}{}^{-1} \, 99\%$  & 4.87 & 5.41 & 4.89 \\
   & $6^+_{  1}$ & $\nu\, 2d_{ 5/2} \, 1g_{ 9/2}{}^{-1} \, 100\%$ &      & 5.40 & 4.88
   & $7^+_{  1}$ & $\nu\, 2d_{ 5/2} \, 1g_{ 9/2}{}^{-1} \, 100\%$ &      & 5.21 & 4.75 \\[\medskipamount]
 h & $2^+_{  2}$ & $\nu\, 1g_{ 7/2} \, 1g_{ 9/2}{}^{-1} \, 97\%$  &      & 8.58 & 8.07
                                                                        & & & & \\
\end{tabular}
\end{ruledtabular}
\end{table*}
\newpage
\bibliographystyle{apsrev4-1}
\bibliography{TTT}

%merlin.mbs apsrev4-1.bst 2010-07-25 4.21a (PWD, AO, DPC) hacked
%Control: key (0)
%Control: author (72) initials jnrlst
%Control: editor formatted (1) identically to author
%Control: production of article title (-1) disabled
%Control: page (0) single
%Control: year (1) truncated
%Control: production of eprint (0) enabled
\begin{thebibliography}{41}%
\makeatletter
\providecommand \@ifxundefined [1]{%
 \@ifx{#1\undefined}
}%
\providecommand \@ifnum [1]{%
 \ifnum #1\expandafter \@firstoftwo
 \else \expandafter \@secondoftwo
 \fi
}%
\providecommand \@ifx [1]{%
 \ifx #1\expandafter \@firstoftwo
 \else \expandafter \@secondoftwo
 \fi
}%
\providecommand \natexlab [1]{#1}%
\providecommand \enquote  [1]{``#1''}%
\providecommand \bibnamefont  [1]{#1}%
\providecommand \bibfnamefont [1]{#1}%
\providecommand \citenamefont [1]{#1}%
\providecommand \href@noop [0]{\@secondoftwo}%
\providecommand \href [0]{\begingroup \@sanitize@url \@href}%
\providecommand \@href[1]{\@@startlink{#1}\@@href}%
\providecommand \@@href[1]{\endgroup#1\@@endlink}%
\providecommand \@sanitize@url [0]{\catcode `\\12\catcode `\$12\catcode
  `\&12\catcode `\#12\catcode `\^12\catcode `\_12\catcode `\%12\relax}%
\providecommand \@@startlink[1]{}%
\providecommand \@@endlink[0]{}%
\providecommand \url  [0]{\begingroup\@sanitize@url \@url }%
\providecommand \@url [1]{\endgroup\@href {#1}{\urlprefix }}%
\providecommand \urlprefix  [0]{URL }%
\providecommand \Eprint [0]{\href }%
\providecommand \doibase [0]{http://dx.doi.org/}%
\providecommand \selectlanguage [0]{\@gobble}%
\providecommand \bibinfo  [0]{\@secondoftwo}%
\providecommand \bibfield  [0]{\@secondoftwo}%
\providecommand \translation [1]{[#1]}%
\providecommand \BibitemOpen [0]{}%
\providecommand \bibitemStop [0]{}%
\providecommand \bibitemNoStop [0]{.\EOS\space}%
\providecommand \EOS [0]{\spacefactor3000\relax}%
\providecommand \BibitemShut  [1]{\csname bibitem#1\endcsname}%
\let\auto@bib@innerbib\@empty
%</preamble>
\bibitem [{\citenamefont {de~Voigt}\ \emph {et~al.}(1983)\citenamefont
  {de~Voigt}, \citenamefont {Dudek},\ and\ \citenamefont
  {Szyma\ifmmode~\acute{n}\else \'{n}\fi{}ski}}]{VoigtRMP83}%
  \BibitemOpen
  \bibfield  {author} {\bibinfo {author} {\bibfnamefont {M.~J.~A.}\
  \bibnamefont {de~Voigt}}, \bibinfo {author} {\bibfnamefont {J.}~\bibnamefont
  {Dudek}}, \ and\ \bibinfo {author} {\bibfnamefont {Z.}~\bibnamefont
  {Szyma\ifmmode~\acute{n}\else \'{n}\fi{}ski}},\ }\href {\doibase
  10.1103/RevModPhys.55.949} {\bibfield  {journal} {\bibinfo  {journal} {Rev.
  Mod. Phys.}\ }\textbf {\bibinfo {volume} {55}},\ \bibinfo {pages} {949}
  (\bibinfo {year} {1983})}\BibitemShut {NoStop}%
\bibitem [{\citenamefont {Saladin}\ \emph {et~al.}(1991)\citenamefont
  {Saladin}, \citenamefont {Sorensen},\ and\ \citenamefont
  {Vincent}}]{SaladinBook91}%
  \BibitemOpen
  \bibfield  {author} {\bibinfo {author} {\bibfnamefont {J.~X.}\ \bibnamefont
  {Saladin}}, \bibinfo {author} {\bibfnamefont {R.~A.}\ \bibnamefont
  {Sorensen}}, \ and\ \bibinfo {author} {\bibfnamefont {C.~M.}\ \bibnamefont
  {Vincent}},\ }\href {\doibase 10.1142/1242} {\emph {\bibinfo {title} {High
  Spin Physics and Gamma-Soft Nuclei}}}\ (\bibinfo  {publisher} {WORLD
  SCIENTIFIC},\ \bibinfo {year} {1991})\ \Eprint
  {http://arxiv.org/abs/https://www.worldscientific.com/doi/pdf/10.1142/1242}
  {https://www.worldscientific.com/doi/pdf/10.1142/1242} \BibitemShut {NoStop}%
\bibitem [{\citenamefont {Ward}\ and\ \citenamefont {Fallon}(2001)}]{Ward2001}%
  \BibitemOpen
  \bibfield  {author} {\bibinfo {author} {\bibfnamefont {D.}~\bibnamefont
  {Ward}}\ and\ \bibinfo {author} {\bibfnamefont {P.}~\bibnamefont {Fallon}},\
  }\enquote {\bibinfo {title} {High spin properties of atomic nuclei},}\ in\
  \href {\doibase 10.1007/0-306-47915-X_3} {\emph {\bibinfo {booktitle}
  {Advances in Nuclear Physics}}},\ \bibinfo {editor} {edited by\ \bibinfo
  {editor} {\bibfnamefont {J.~W.}\ \bibnamefont {Negele}}\ and\ \bibinfo
  {editor} {\bibfnamefont {E.~W.}\ \bibnamefont {Vogt}}}\ (\bibinfo
  {publisher} {Springer US},\ \bibinfo {address} {Boston, MA},\ \bibinfo {year}
  {2001})\ pp.\ \bibinfo {pages} {167--291}\BibitemShut {NoStop}%
\bibitem [{\citenamefont {Meng}\ \emph {et~al.}(2016)\citenamefont {Meng},
  \citenamefont {Zhang},\ and\ \citenamefont {Zhao}}]{Meng_2016}%
  \BibitemOpen
  \bibfield  {author} {\bibinfo {author} {\bibfnamefont {J.}~\bibnamefont
  {Meng}}, \bibinfo {author} {\bibfnamefont {S.~Q.}\ \bibnamefont {Zhang}}, \
  and\ \bibinfo {author} {\bibfnamefont {P.~W.}\ \bibnamefont {Zhao}},\ }in\
  \href@noop {} {\emph {\bibinfo {booktitle} {Relativistic Density Functional
  for Nuclear Structure}}},\ Vol.\ \bibinfo {volume} {International Review of
  Nuclear Physics -- Vol. 10},\ \bibinfo {editor} {edited by\ \bibinfo {editor}
  {\bibfnamefont {J.}~\bibnamefont {Meng}}}\ (\bibinfo  {publisher} {World
  Scientific Publishing Co. Pte. Ltd.},\ \bibinfo {year} {2016})\ pp.\ \bibinfo
  {pages} {355--412}\BibitemShut {NoStop}%
\bibitem [{\citenamefont {Frauendorf}(2018)}]{Frauendorf_2018}%
  \BibitemOpen
  \bibfield  {author} {\bibinfo {author} {\bibfnamefont {S.}~\bibnamefont
  {Frauendorf}},\ }\href {\doibase 10.1088/1402-4896/aaa2e9} {\bibfield
  {journal} {\bibinfo  {journal} {Physica Scripta}\ }\textbf {\bibinfo {volume}
  {93}},\ \bibinfo {pages} {043003} (\bibinfo {year} {2018})}\BibitemShut
  {NoStop}%
\bibitem [{\citenamefont {Petrache}\ \emph {et~al.}(2019)\citenamefont
  {Petrache}, \citenamefont {Walker}, \citenamefont {Guo}, \citenamefont
  {Chen}, \citenamefont {Frauendorf}, \citenamefont {Liu}, \citenamefont
  {Wyss}, \citenamefont {Mengoni}, \citenamefont {Qiang}, \citenamefont
  {Astier}, \citenamefont {Dupont}, \citenamefont {Li}, \citenamefont {Lv},
  \citenamefont {Zheng}, \citenamefont {Bazzacco}, \citenamefont {Boso},
  \citenamefont {Goasduff}, \citenamefont {Recchia}, \citenamefont {Testov},
  \citenamefont {Galtarossa}, \citenamefont {Jaworski}, \citenamefont {Napoli},
  \citenamefont {Riccetto}, \citenamefont {Siciliano}, \citenamefont
  {Valiente-Dobon}, \citenamefont {Liu}, \citenamefont {Zhou}, \citenamefont
  {Wang}, \citenamefont {Andreoiu}, \citenamefont {Garcia}, \citenamefont
  {Ortner}, \citenamefont {Whitmore}, \citenamefont {Back}, \citenamefont
  {Cederwall}, \citenamefont {Lawrie}, \citenamefont {Kuti}, \citenamefont
  {Sohler}, \citenamefont {Timar}, \citenamefont {Marchlewski}, \citenamefont
  {Srebrny},\ and\ \citenamefont {Tucholski}}]{Petrache_2019}%
  \BibitemOpen
  \bibfield  {author} {\bibinfo {author} {\bibfnamefont {C.}~\bibnamefont
  {Petrache}}, \bibinfo {author} {\bibfnamefont {P.}~\bibnamefont {Walker}},
  \bibinfo {author} {\bibfnamefont {S.}~\bibnamefont {Guo}}, \bibinfo {author}
  {\bibfnamefont {Q.}~\bibnamefont {Chen}}, \bibinfo {author} {\bibfnamefont
  {S.}~\bibnamefont {Frauendorf}}, \bibinfo {author} {\bibfnamefont
  {Y.}~\bibnamefont {Liu}}, \bibinfo {author} {\bibfnamefont {R.}~\bibnamefont
  {Wyss}}, \bibinfo {author} {\bibfnamefont {D.}~\bibnamefont {Mengoni}},
  \bibinfo {author} {\bibfnamefont {Y.}~\bibnamefont {Qiang}}, \bibinfo
  {author} {\bibfnamefont {A.}~\bibnamefont {Astier}}, \bibinfo {author}
  {\bibfnamefont {E.}~\bibnamefont {Dupont}}, \bibinfo {author} {\bibfnamefont
  {R.}~\bibnamefont {Li}}, \bibinfo {author} {\bibfnamefont {B.}~\bibnamefont
  {Lv}}, \bibinfo {author} {\bibfnamefont {K.}~\bibnamefont {Zheng}}, \bibinfo
  {author} {\bibfnamefont {D.}~\bibnamefont {Bazzacco}}, \bibinfo {author}
  {\bibfnamefont {A.}~\bibnamefont {Boso}}, \bibinfo {author} {\bibfnamefont
  {A.}~\bibnamefont {Goasduff}}, \bibinfo {author} {\bibfnamefont
  {F.}~\bibnamefont {Recchia}}, \bibinfo {author} {\bibfnamefont
  {D.}~\bibnamefont {Testov}}, \bibinfo {author} {\bibfnamefont
  {F.}~\bibnamefont {Galtarossa}}, \bibinfo {author} {\bibfnamefont
  {G.}~\bibnamefont {Jaworski}}, \bibinfo {author} {\bibfnamefont
  {D.}~\bibnamefont {Napoli}}, \bibinfo {author} {\bibfnamefont
  {S.}~\bibnamefont {Riccetto}}, \bibinfo {author} {\bibfnamefont
  {M.}~\bibnamefont {Siciliano}}, \bibinfo {author} {\bibfnamefont
  {J.}~\bibnamefont {Valiente-Dobon}}, \bibinfo {author} {\bibfnamefont
  {M.}~\bibnamefont {Liu}}, \bibinfo {author} {\bibfnamefont {X.}~\bibnamefont
  {Zhou}}, \bibinfo {author} {\bibfnamefont {J.}~\bibnamefont {Wang}}, \bibinfo
  {author} {\bibfnamefont {C.}~\bibnamefont {Andreoiu}}, \bibinfo {author}
  {\bibfnamefont {F.}~\bibnamefont {Garcia}}, \bibinfo {author} {\bibfnamefont
  {K.}~\bibnamefont {Ortner}}, \bibinfo {author} {\bibfnamefont
  {K.}~\bibnamefont {Whitmore}}, \bibinfo {author} {\bibfnamefont
  {T.}~\bibnamefont {Back}}, \bibinfo {author} {\bibfnamefont {B.}~\bibnamefont
  {Cederwall}}, \bibinfo {author} {\bibfnamefont {E.}~\bibnamefont {Lawrie}},
  \bibinfo {author} {\bibfnamefont {I.}~\bibnamefont {Kuti}}, \bibinfo {author}
  {\bibfnamefont {D.}~\bibnamefont {Sohler}}, \bibinfo {author} {\bibfnamefont
  {J.}~\bibnamefont {Timar}}, \bibinfo {author} {\bibfnamefont
  {T.}~\bibnamefont {Marchlewski}}, \bibinfo {author} {\bibfnamefont
  {J.}~\bibnamefont {Srebrny}}, \ and\ \bibinfo {author} {\bibfnamefont
  {A.}~\bibnamefont {Tucholski}},\ }\href {\doibase
  https://doi.org/10.1016/j.physletb.2019.06.040} {\bibfield  {journal}
  {\bibinfo  {journal} {Physics Letters B}\ }\textbf {\bibinfo {volume}
  {795}},\ \bibinfo {pages} {241} (\bibinfo {year} {2019})}\BibitemShut
  {NoStop}%
\bibitem [{\citenamefont {Kumar}\ and\ \citenamefont
  {Srivastava}(2020)}]{Kumar_2020}%
  \BibitemOpen
  \bibfield  {author} {\bibinfo {author} {\bibfnamefont {V.}~\bibnamefont
  {Kumar}}\ and\ \bibinfo {author} {\bibfnamefont {P.~C.}\ \bibnamefont
  {Srivastava}},\ }\href {\doibase
  https://doi.org/10.1016/j.nuclphysa.2020.121989} {\bibfield  {journal}
  {\bibinfo  {journal} {Nuclear Physics A}\ }\textbf {\bibinfo {volume}
  {1002}},\ \bibinfo {pages} {121989} (\bibinfo {year} {2020})}\BibitemShut
  {NoStop}%
\bibitem [{\citenamefont {Afanasjev}(2022)}]{Afanasjev_2022}%
  \BibitemOpen
  \bibfield  {author} {\bibinfo {author} {\bibfnamefont {A.~V.}\ \bibnamefont
  {Afanasjev}},\ }\enquote {\bibinfo {title} {Model for independent particle
  motion},}\ in\ \href {\doibase 10.1007/978-981-15-8818-1_10-1} {\emph
  {\bibinfo {booktitle} {Handbook of Nuclear Physics}}},\ \bibinfo {editor}
  {edited by\ \bibinfo {editor} {\bibfnamefont {I.}~\bibnamefont {Tanihata}},
  \bibinfo {editor} {\bibfnamefont {H.}~\bibnamefont {Toki}}, \ and\ \bibinfo
  {editor} {\bibfnamefont {T.}~\bibnamefont {Kajino}}}\ (\bibinfo  {publisher}
  {Springer Nature Singapore},\ \bibinfo {address} {Singapore},\ \bibinfo
  {year} {2022})\ pp.\ \bibinfo {pages} {1--40}\BibitemShut {NoStop}%
\bibitem [{\citenamefont {Yoshida}(2022)}]{Yoshida_2022}%
  \BibitemOpen
  \bibfield  {author} {\bibinfo {author} {\bibfnamefont {K.}~\bibnamefont
  {Yoshida}},\ }\href {\doibase 10.1103/PhysRevC.105.024318} {\bibfield
  {journal} {\bibinfo  {journal} {Phys. Rev. C}\ }\textbf {\bibinfo {volume}
  {105}},\ \bibinfo {pages} {024318} (\bibinfo {year} {2022})}\BibitemShut
  {NoStop}%
\bibitem [{\citenamefont {Rudolph}\ \emph {et~al.}(1999)\citenamefont
  {Rudolph}, \citenamefont {Baktash}, \citenamefont {Brinkman}, \citenamefont
  {Caurier}, \citenamefont {Dean}, \citenamefont {Devlin}, \citenamefont
  {Dobaczewski}, \citenamefont {Heenen}, \citenamefont {Jin}, \citenamefont
  {LaFosse}, \citenamefont {Nazarewicz}, \citenamefont {Nowacki}, \citenamefont
  {Poves}, \citenamefont {Riedinger}, \citenamefont {Sarantites}, \citenamefont
  {Satu\l{}a},\ and\ \citenamefont {Yu}}]{Rudolph_1999}%
  \BibitemOpen
  \bibfield  {author} {\bibinfo {author} {\bibfnamefont {D.}~\bibnamefont
  {Rudolph}}, \bibinfo {author} {\bibfnamefont {C.}~\bibnamefont {Baktash}},
  \bibinfo {author} {\bibfnamefont {M.~J.}\ \bibnamefont {Brinkman}}, \bibinfo
  {author} {\bibfnamefont {E.}~\bibnamefont {Caurier}}, \bibinfo {author}
  {\bibfnamefont {D.~J.}\ \bibnamefont {Dean}}, \bibinfo {author}
  {\bibfnamefont {M.}~\bibnamefont {Devlin}}, \bibinfo {author} {\bibfnamefont
  {J.}~\bibnamefont {Dobaczewski}}, \bibinfo {author} {\bibfnamefont {P.-H.}\
  \bibnamefont {Heenen}}, \bibinfo {author} {\bibfnamefont {H.-Q.}\
  \bibnamefont {Jin}}, \bibinfo {author} {\bibfnamefont {D.~R.}\ \bibnamefont
  {LaFosse}}, \bibinfo {author} {\bibfnamefont {W.}~\bibnamefont {Nazarewicz}},
  \bibinfo {author} {\bibfnamefont {F.}~\bibnamefont {Nowacki}}, \bibinfo
  {author} {\bibfnamefont {A.}~\bibnamefont {Poves}}, \bibinfo {author}
  {\bibfnamefont {L.~L.}\ \bibnamefont {Riedinger}}, \bibinfo {author}
  {\bibfnamefont {D.~G.}\ \bibnamefont {Sarantites}}, \bibinfo {author}
  {\bibfnamefont {W.}~\bibnamefont {Satu\l{}a}}, \ and\ \bibinfo {author}
  {\bibfnamefont {C.-H.}\ \bibnamefont {Yu}},\ }\href {\doibase
  10.1103/PhysRevLett.82.3763} {\bibfield  {journal} {\bibinfo  {journal}
  {Phys. Rev. Lett.}\ }\textbf {\bibinfo {volume} {82}},\ \bibinfo {pages}
  {3763} (\bibinfo {year} {1999})}\BibitemShut {NoStop}%
\bibitem [{\citenamefont {Ideguchi}\ \emph {et~al.}(2001)\citenamefont
  {Ideguchi}, \citenamefont {Sarantites}, \citenamefont {Reviol}, \citenamefont
  {Afanasjev}, \citenamefont {Devlin}, \citenamefont {Baktash}, \citenamefont
  {Janssens}, \citenamefont {Rudolph}, \citenamefont {Axelsson}, \citenamefont
  {Carpenter}, \citenamefont {Galindo-Uribarri}, \citenamefont {LaFosse},
  \citenamefont {Lauritsen}, \citenamefont {Lerma}, \citenamefont {Lister},
  \citenamefont {Reiter}, \citenamefont {Seweryniak}, \citenamefont
  {Weiszflog},\ and\ \citenamefont {Wilson}}]{Ideguchi_2001}%
  \BibitemOpen
  \bibfield  {author} {\bibinfo {author} {\bibfnamefont {E.}~\bibnamefont
  {Ideguchi}}, \bibinfo {author} {\bibfnamefont {D.~G.}\ \bibnamefont
  {Sarantites}}, \bibinfo {author} {\bibfnamefont {W.}~\bibnamefont {Reviol}},
  \bibinfo {author} {\bibfnamefont {A.~V.}\ \bibnamefont {Afanasjev}}, \bibinfo
  {author} {\bibfnamefont {M.}~\bibnamefont {Devlin}}, \bibinfo {author}
  {\bibfnamefont {C.}~\bibnamefont {Baktash}}, \bibinfo {author} {\bibfnamefont
  {R.~V.~F.}\ \bibnamefont {Janssens}}, \bibinfo {author} {\bibfnamefont
  {D.}~\bibnamefont {Rudolph}}, \bibinfo {author} {\bibfnamefont
  {A.}~\bibnamefont {Axelsson}}, \bibinfo {author} {\bibfnamefont {M.~P.}\
  \bibnamefont {Carpenter}}, \bibinfo {author} {\bibfnamefont {A.}~\bibnamefont
  {Galindo-Uribarri}}, \bibinfo {author} {\bibfnamefont {D.~R.}\ \bibnamefont
  {LaFosse}}, \bibinfo {author} {\bibfnamefont {T.}~\bibnamefont {Lauritsen}},
  \bibinfo {author} {\bibfnamefont {F.}~\bibnamefont {Lerma}}, \bibinfo
  {author} {\bibfnamefont {C.~J.}\ \bibnamefont {Lister}}, \bibinfo {author}
  {\bibfnamefont {P.}~\bibnamefont {Reiter}}, \bibinfo {author} {\bibfnamefont
  {D.}~\bibnamefont {Seweryniak}}, \bibinfo {author} {\bibfnamefont
  {M.}~\bibnamefont {Weiszflog}}, \ and\ \bibinfo {author} {\bibfnamefont
  {J.~N.}\ \bibnamefont {Wilson}},\ }\href {\doibase
  10.1103/PhysRevLett.87.222501} {\bibfield  {journal} {\bibinfo  {journal}
  {Phys. Rev. Lett.}\ }\textbf {\bibinfo {volume} {87}},\ \bibinfo {pages}
  {222501} (\bibinfo {year} {2001})}\BibitemShut {NoStop}%
\bibitem [{\citenamefont {Inakura}\ \emph {et~al.}(2002)\citenamefont
  {Inakura}, \citenamefont {Mizutori}, \citenamefont {Yamagami},\ and\
  \citenamefont {Matsuyanagi}}]{Inakura_2002}%
  \BibitemOpen
  \bibfield  {author} {\bibinfo {author} {\bibfnamefont {T.}~\bibnamefont
  {Inakura}}, \bibinfo {author} {\bibfnamefont {S.}~\bibnamefont {Mizutori}},
  \bibinfo {author} {\bibfnamefont {M.}~\bibnamefont {Yamagami}}, \ and\
  \bibinfo {author} {\bibfnamefont {K.}~\bibnamefont {Matsuyanagi}},\ }\href
  {\doibase https://doi.org/10.1016/S0375-9474(02)01164-8} {\bibfield
  {journal} {\bibinfo  {journal} {Nuclear Physics A}\ }\textbf {\bibinfo
  {volume} {710}},\ \bibinfo {pages} {261} (\bibinfo {year}
  {2002})}\BibitemShut {NoStop}%
\bibitem [{\citenamefont {Caurier}\ \emph {et~al.}(2007)\citenamefont
  {Caurier}, \citenamefont {Men\'endez}, \citenamefont {Nowacki},\ and\
  \citenamefont {Poves}}]{Caurier_2007}%
  \BibitemOpen
  \bibfield  {author} {\bibinfo {author} {\bibfnamefont {E.}~\bibnamefont
  {Caurier}}, \bibinfo {author} {\bibfnamefont {J.}~\bibnamefont {Men\'endez}},
  \bibinfo {author} {\bibfnamefont {F.}~\bibnamefont {Nowacki}}, \ and\
  \bibinfo {author} {\bibfnamefont {A.}~\bibnamefont {Poves}},\ }\href
  {\doibase 10.1103/PhysRevC.75.054317} {\bibfield  {journal} {\bibinfo
  {journal} {Phys. Rev. C}\ }\textbf {\bibinfo {volume} {75}},\ \bibinfo
  {pages} {054317} (\bibinfo {year} {2007})}\BibitemShut {NoStop}%
\bibitem [{\citenamefont {Chiba}\ and\ \citenamefont
  {Kimura}(2014)}]{Chiba_2014}%
  \BibitemOpen
  \bibfield  {author} {\bibinfo {author} {\bibfnamefont {Y.}~\bibnamefont
  {Chiba}}\ and\ \bibinfo {author} {\bibfnamefont {M.}~\bibnamefont {Kimura}},\
  }\href {\doibase 10.1103/PhysRevC.89.054313} {\bibfield  {journal} {\bibinfo
  {journal} {Phys. Rev. C}\ }\textbf {\bibinfo {volume} {89}},\ \bibinfo
  {pages} {054313} (\bibinfo {year} {2014})}\BibitemShut {NoStop}%
\bibitem [{\citenamefont {Broda}\ \emph {et~al.}(2017)\citenamefont {Broda},
  \citenamefont {Janssens}, \citenamefont {Iskra}, \citenamefont {Wrzesinski},
  \citenamefont {Fornal}, \citenamefont {Carpenter}, \citenamefont {Chiara},
  \citenamefont {Cieplicka-Ory\ifmmode~\acute{n}\else \'{n}\fi{}czak},
  \citenamefont {Hoffman}, \citenamefont {Kondev}, \citenamefont {Kr\'olas},
  \citenamefont {Lauritsen}, \citenamefont {Podolyak}, \citenamefont
  {Seweryniak}, \citenamefont {Shand}, \citenamefont {Szpak}, \citenamefont
  {Walters}, \citenamefont {Zhu},\ and\ \citenamefont {Brown}}]{Broda_2017}%
  \BibitemOpen
  \bibfield  {author} {\bibinfo {author} {\bibfnamefont {R.}~\bibnamefont
  {Broda}}, \bibinfo {author} {\bibfnamefont {R.~V.~F.}\ \bibnamefont
  {Janssens}}, \bibinfo {author} {\bibfnamefont {L.~W.}\ \bibnamefont {Iskra}},
  \bibinfo {author} {\bibfnamefont {J.}~\bibnamefont {Wrzesinski}}, \bibinfo
  {author} {\bibfnamefont {B.}~\bibnamefont {Fornal}}, \bibinfo {author}
  {\bibfnamefont {M.~P.}\ \bibnamefont {Carpenter}}, \bibinfo {author}
  {\bibfnamefont {C.~J.}\ \bibnamefont {Chiara}}, \bibinfo {author}
  {\bibfnamefont {N.}~\bibnamefont {Cieplicka-Ory\ifmmode~\acute{n}\else
  \'{n}\fi{}czak}}, \bibinfo {author} {\bibfnamefont {C.~R.}\ \bibnamefont
  {Hoffman}}, \bibinfo {author} {\bibfnamefont {F.~G.}\ \bibnamefont {Kondev}},
  \bibinfo {author} {\bibfnamefont {W.}~\bibnamefont {Kr\'olas}}, \bibinfo
  {author} {\bibfnamefont {T.}~\bibnamefont {Lauritsen}}, \bibinfo {author}
  {\bibfnamefont {Z.}~\bibnamefont {Podolyak}}, \bibinfo {author}
  {\bibfnamefont {D.}~\bibnamefont {Seweryniak}}, \bibinfo {author}
  {\bibfnamefont {C.~M.}\ \bibnamefont {Shand}}, \bibinfo {author}
  {\bibfnamefont {B.}~\bibnamefont {Szpak}}, \bibinfo {author} {\bibfnamefont
  {W.~B.}\ \bibnamefont {Walters}}, \bibinfo {author} {\bibfnamefont
  {S.}~\bibnamefont {Zhu}}, \ and\ \bibinfo {author} {\bibfnamefont {B.~A.}\
  \bibnamefont {Brown}},\ }\href {\doibase 10.1103/PhysRevC.95.064308}
  {\bibfield  {journal} {\bibinfo  {journal} {Phys. Rev. C}\ }\textbf {\bibinfo
  {volume} {95}},\ \bibinfo {pages} {064308} (\bibinfo {year}
  {2017})}\BibitemShut {NoStop}%
\bibitem [{\citenamefont {Sakai}\ \emph {et~al.}(2020)\citenamefont {Sakai},
  \citenamefont {Yoshida},\ and\ \citenamefont {Matsuo}}]{Sakai_2020}%
  \BibitemOpen
  \bibfield  {author} {\bibinfo {author} {\bibfnamefont {S.}~\bibnamefont
  {Sakai}}, \bibinfo {author} {\bibfnamefont {K.}~\bibnamefont {Yoshida}}, \
  and\ \bibinfo {author} {\bibfnamefont {M.}~\bibnamefont {Matsuo}},\ }\href
  {\doibase 10.1093/ptep/ptaa071} {\bibfield  {journal} {\bibinfo  {journal}
  {Progress of Theoretical and Experimental Physics}\ }\textbf {\bibinfo
  {volume} {2020}} (\bibinfo {year} {2020}),\ 10.1093/ptep/ptaa071},\ \bibinfo
  {note} {063D02},\ \Eprint
  {http://arxiv.org/abs/https://academic.oup.com/ptep/article-pdf/2020/6/063D02/33424844/ptaa071.pdf}
  {https://academic.oup.com/ptep/article-pdf/2020/6/063D02/33424844/ptaa071.pdf}
  \BibitemShut {NoStop}%
\bibitem [{\citenamefont {Wang}\ \emph {et~al.}(2020)\citenamefont {Wang},
  \citenamefont {Shi},\ and\ \citenamefont {Xia}}]{Wang_2020}%
  \BibitemOpen
  \bibfield  {author} {\bibinfo {author} {\bibfnamefont {F.}~\bibnamefont
  {Wang}}, \bibinfo {author} {\bibfnamefont {Z.}~\bibnamefont {Shi}}, \ and\
  \bibinfo {author} {\bibfnamefont {X.~W.}\ \bibnamefont {Xia}},\ }\href
  {\doibase 10.1103/PhysRevC.102.014321} {\bibfield  {journal} {\bibinfo
  {journal} {Phys. Rev. C}\ }\textbf {\bibinfo {volume} {102}},\ \bibinfo
  {pages} {014321} (\bibinfo {year} {2020})}\BibitemShut {NoStop}%
\bibitem [{\citenamefont {Dey}\ \emph {et~al.}(2022)\citenamefont {Dey},
  \citenamefont {Negi}, \citenamefont {Palit}, \citenamefont {Srivastava},
  \citenamefont {Laskar}, \citenamefont {Das}, \citenamefont {Babra},
  \citenamefont {Bhattacharya}, \citenamefont {Das}, \citenamefont {Devi},
  \citenamefont {Gala}, \citenamefont {Garg}, \citenamefont {Ghugre},
  \citenamefont {Ideguchi}, \citenamefont {Kumar}, \citenamefont {Kundu},
  \citenamefont {Mukherjee}, \citenamefont {Muralithar}, \citenamefont {Nag},
  \citenamefont {Nandi}, \citenamefont {Neelam}, \citenamefont {Raja},
  \citenamefont {Raut}, \citenamefont {Santra}, \citenamefont {Sharma},
  \citenamefont {Sihotra}, \citenamefont {Singh}, \citenamefont {Singh},\ and\
  \citenamefont {Trivedi}}]{Dey_2022}%
  \BibitemOpen
  \bibfield  {author} {\bibinfo {author} {\bibfnamefont {P.}~\bibnamefont
  {Dey}}, \bibinfo {author} {\bibfnamefont {D.}~\bibnamefont {Negi}}, \bibinfo
  {author} {\bibfnamefont {R.}~\bibnamefont {Palit}}, \bibinfo {author}
  {\bibfnamefont {P.~C.}\ \bibnamefont {Srivastava}}, \bibinfo {author}
  {\bibfnamefont {M.~S.~R.}\ \bibnamefont {Laskar}}, \bibinfo {author}
  {\bibfnamefont {B.}~\bibnamefont {Das}}, \bibinfo {author} {\bibfnamefont
  {F.~S.}\ \bibnamefont {Babra}}, \bibinfo {author} {\bibfnamefont
  {S.}~\bibnamefont {Bhattacharya}}, \bibinfo {author} {\bibfnamefont
  {B.}~\bibnamefont {Das}}, \bibinfo {author} {\bibfnamefont {K.~R.}\
  \bibnamefont {Devi}}, \bibinfo {author} {\bibfnamefont {R.}~\bibnamefont
  {Gala}}, \bibinfo {author} {\bibfnamefont {U.}~\bibnamefont {Garg}}, \bibinfo
  {author} {\bibfnamefont {S.~S.}\ \bibnamefont {Ghugre}}, \bibinfo {author}
  {\bibfnamefont {E.}~\bibnamefont {Ideguchi}}, \bibinfo {author}
  {\bibfnamefont {S.}~\bibnamefont {Kumar}}, \bibinfo {author} {\bibfnamefont
  {A.}~\bibnamefont {Kundu}}, \bibinfo {author} {\bibfnamefont
  {G.}~\bibnamefont {Mukherjee}}, \bibinfo {author} {\bibfnamefont
  {S.}~\bibnamefont {Muralithar}}, \bibinfo {author} {\bibfnamefont
  {S.}~\bibnamefont {Nag}}, \bibinfo {author} {\bibfnamefont {S.}~\bibnamefont
  {Nandi}}, \bibinfo {author} {\bibnamefont {Neelam}}, \bibinfo {author}
  {\bibfnamefont {M.~K.}\ \bibnamefont {Raja}}, \bibinfo {author}
  {\bibfnamefont {R.}~\bibnamefont {Raut}}, \bibinfo {author} {\bibfnamefont
  {R.}~\bibnamefont {Santra}}, \bibinfo {author} {\bibfnamefont
  {A.}~\bibnamefont {Sharma}}, \bibinfo {author} {\bibfnamefont
  {S.}~\bibnamefont {Sihotra}}, \bibinfo {author} {\bibfnamefont {A.~K.}\
  \bibnamefont {Singh}}, \bibinfo {author} {\bibfnamefont {R.~P.}\ \bibnamefont
  {Singh}}, \ and\ \bibinfo {author} {\bibfnamefont {T.}~\bibnamefont
  {Trivedi}},\ }\href {\doibase 10.1103/PhysRevC.105.044307} {\bibfield
  {journal} {\bibinfo  {journal} {Phys. Rev. C}\ }\textbf {\bibinfo {volume}
  {105}},\ \bibinfo {pages} {044307} (\bibinfo {year} {2022})}\BibitemShut
  {NoStop}%
\bibitem [{\citenamefont {Warburton}\ \emph {et~al.}(1985)\citenamefont
  {Warburton}, \citenamefont {Olness}, \citenamefont {Lister}, \citenamefont
  {Zurm\"uhle},\ and\ \citenamefont {Becker}}]{Warburton_1985}%
  \BibitemOpen
  \bibfield  {author} {\bibinfo {author} {\bibfnamefont {E.~K.}\ \bibnamefont
  {Warburton}}, \bibinfo {author} {\bibfnamefont {J.~W.}\ \bibnamefont
  {Olness}}, \bibinfo {author} {\bibfnamefont {C.~J.}\ \bibnamefont {Lister}},
  \bibinfo {author} {\bibfnamefont {R.~W.}\ \bibnamefont {Zurm\"uhle}}, \ and\
  \bibinfo {author} {\bibfnamefont {J.~A.}\ \bibnamefont {Becker}},\ }\href
  {\doibase 10.1103/PhysRevC.31.1184} {\bibfield  {journal} {\bibinfo
  {journal} {Phys. Rev. C}\ }\textbf {\bibinfo {volume} {31}},\ \bibinfo
  {pages} {1184} (\bibinfo {year} {1985})}\BibitemShut {NoStop}%
\bibitem [{\citenamefont {Wang}\ \emph {et~al.}(2021)\citenamefont {Wang},
  \citenamefont {Ma}, \citenamefont {Liu},\ and\ \citenamefont
  {Lu}}]{Wang_2021}%
  \BibitemOpen
  \bibfield  {author} {\bibinfo {author} {\bibfnamefont {H.}~\bibnamefont
  {Wang}}, \bibinfo {author} {\bibfnamefont {K.-Y.}\ \bibnamefont {Ma}},
  \bibinfo {author} {\bibfnamefont {S.-Y.}\ \bibnamefont {Liu}}, \ and\
  \bibinfo {author} {\bibfnamefont {J.-B.}\ \bibnamefont {Lu}},\ }\href
  {\doibase 10.1088/1674-1137/ac0fd2} {\bibfield  {journal} {\bibinfo
  {journal} {Chinese Physics C}\ }\textbf {\bibinfo {volume} {45}},\ \bibinfo
  {pages} {094106} (\bibinfo {year} {2021})}\BibitemShut {NoStop}%
\bibitem [{\citenamefont {Chen}(2017)}]{Chen_2017}%
  \BibitemOpen
  \bibfield  {author} {\bibinfo {author} {\bibfnamefont {J.}~\bibnamefont
  {Chen}},\ }\href {\doibase http://dx.doi.org/10.1016/j.nds.2017.02.001}
  {\bibfield  {journal} {\bibinfo  {journal} {Nuclear Data Sheets}\ }\textbf
  {\bibinfo {volume} {140}},\ \bibinfo {pages} {1 } (\bibinfo {year}
  {2017})}\BibitemShut {NoStop}%
\bibitem [{\citenamefont {Oi}(2007)}]{Oi_2007}%
  \BibitemOpen
  \bibfield  {author} {\bibinfo {author} {\bibfnamefont {M.}~\bibnamefont
  {Oi}},\ }\href {\doibase 10.1103/PhysRevC.76.044308} {\bibfield  {journal}
  {\bibinfo  {journal} {Phys. Rev. C}\ }\textbf {\bibinfo {volume} {76}},\
  \bibinfo {pages} {044308} (\bibinfo {year} {2007})}\BibitemShut {NoStop}%
\bibitem [{\citenamefont {Lyutorovich}\ \emph {et~al.}(2022)\citenamefont
  {Lyutorovich}, \citenamefont {Tselyaev}, \citenamefont {Speth}, \citenamefont
  {Martinez-Pinedo}, \citenamefont {Langanke},\ and\ \citenamefont
  {Reinhard}}]{Lyutorovich_2022}%
  \BibitemOpen
  \bibfield  {author} {\bibinfo {author} {\bibfnamefont {N.}~\bibnamefont
  {Lyutorovich}}, \bibinfo {author} {\bibfnamefont {V.}~\bibnamefont
  {Tselyaev}}, \bibinfo {author} {\bibfnamefont {J.}~\bibnamefont {Speth}},
  \bibinfo {author} {\bibfnamefont {G.}~\bibnamefont {Martinez-Pinedo}},
  \bibinfo {author} {\bibfnamefont {K.}~\bibnamefont {Langanke}}, \ and\
  \bibinfo {author} {\bibfnamefont {P.-G.}\ \bibnamefont {Reinhard}},\ }\href
  {\doibase 10.1103/PhysRevC.105.014327} {\bibfield  {journal} {\bibinfo
  {journal} {Phys. Rev. C}\ }\textbf {\bibinfo {volume} {105}},\ \bibinfo
  {pages} {014327} (\bibinfo {year} {2022})}\BibitemShut {NoStop}%
\bibitem [{\citenamefont {Tselyaev}\ \emph {et~al.}(2018)\citenamefont
  {Tselyaev}, \citenamefont {Lyutorovich}, \citenamefont {Speth},\ and\
  \citenamefont {Reinhard}}]{Tselyaev_2018}%
  \BibitemOpen
  \bibfield  {author} {\bibinfo {author} {\bibfnamefont {V.}~\bibnamefont
  {Tselyaev}}, \bibinfo {author} {\bibfnamefont {N.}~\bibnamefont
  {Lyutorovich}}, \bibinfo {author} {\bibfnamefont {J.}~\bibnamefont {Speth}},
  \ and\ \bibinfo {author} {\bibfnamefont {P.-G.}\ \bibnamefont {Reinhard}},\
  }\href {\doibase 10.1103/PhysRevC.97.044308} {\bibfield  {journal} {\bibinfo
  {journal} {Phys. Rev. C}\ }\textbf {\bibinfo {volume} {97}},\ \bibinfo
  {pages} {044308} (\bibinfo {year} {2018})}\BibitemShut {NoStop}%
\bibitem [{\citenamefont {Vdovin}\ and\ \citenamefont
  {Soloviev}(1983)}]{Vdovin_1983}%
  \BibitemOpen
  \bibfield  {author} {\bibinfo {author} {\bibfnamefont {A.~I.}\ \bibnamefont
  {Vdovin}}\ and\ \bibinfo {author} {\bibfnamefont {V.~G.}\ \bibnamefont
  {Soloviev}},\ }\href@noop {} {\bibfield  {journal} {\bibinfo  {journal} {Sov.
  J. Part. Nucl.}\ }\textbf {\bibinfo {volume} {14}},\ \bibinfo {pages} {99}
  (\bibinfo {year} {1983})}\BibitemShut {NoStop}%
\bibitem [{\citenamefont {Voronov}\ and\ \citenamefont
  {Soloviev}(1983)}]{Voronov_1983}%
  \BibitemOpen
  \bibfield  {author} {\bibinfo {author} {\bibfnamefont {V.~V.}\ \bibnamefont
  {Voronov}}\ and\ \bibinfo {author} {\bibfnamefont {V.~G.}\ \bibnamefont
  {Soloviev}},\ }\href@noop {} {\bibfield  {journal} {\bibinfo  {journal} {Sov.
  J. Part. Nucl.}\ }\textbf {\bibinfo {volume} {14}},\ \bibinfo {pages} {583}
  (\bibinfo {year} {1983})}\BibitemShut {NoStop}%
\bibitem [{\citenamefont {Soloviev}(1992)}]{Soloviev_1992}%
  \BibitemOpen
  \bibfield  {author} {\bibinfo {author} {\bibfnamefont {V.~G.}\ \bibnamefont
  {Soloviev}},\ }\href@noop {} {\emph {\bibinfo {title} {Theory of Atomic
  Nuclei: Quasiparticles and Phonons}}}\ (\bibinfo  {publisher} {Institute of
  Physics},\ \bibinfo {address} {Bristol and Philadelphia},\ \bibinfo {year}
  {1992})\BibitemShut {NoStop}%
\bibitem [{\citenamefont {{C. A. Bertulani}}\ and\ \citenamefont {{V. Yu.
  Ponomarev}}(1999)}]{Bertulani_1999}%
  \BibitemOpen
  \bibfield  {author} {\bibinfo {author} {\bibnamefont {{C. A. Bertulani}}}\
  and\ \bibinfo {author} {\bibnamefont {{V. Yu. Ponomarev}}},\ }\href@noop {}
  {\bibfield  {journal} {\bibinfo  {journal} {Phys. Rep.}\ }\textbf {\bibinfo
  {volume} {321}},\ \bibinfo {pages} {139} (\bibinfo {year}
  {1999})}\BibitemShut {NoStop}%
\bibitem [{\citenamefont {Van~Giai}\ \emph {et~al.}(1998)\citenamefont
  {Van~Giai}, \citenamefont {Stoyanov},\ and\ \citenamefont
  {Voronov}}]{Giai_1998}%
  \BibitemOpen
  \bibfield  {author} {\bibinfo {author} {\bibfnamefont {N.}~\bibnamefont
  {Van~Giai}}, \bibinfo {author} {\bibfnamefont {C.}~\bibnamefont {Stoyanov}},
  \ and\ \bibinfo {author} {\bibfnamefont {V.}~\bibnamefont {Voronov}},\
  }\href@noop {} {\bibfield  {journal} {\bibinfo  {journal} {Phys. Rev. C}\
  }\textbf {\bibinfo {volume} {57}},\ \bibinfo {pages} {1204} (\bibinfo {year}
  {1998})}\BibitemShut {NoStop}%
\bibitem [{\citenamefont {{A. P. Severyukhin}}\ \emph
  {et~al.}(2018)\citenamefont {{A. P. Severyukhin}}, \citenamefont {{S.
  \AA{}berg}}, \citenamefont {{N. N. Arsenyev}},\ and\ \citenamefont {{R. G.
  Nazmitdinov}}}]{Severyukhin_2018prc}%
  \BibitemOpen
  \bibfield  {author} {\bibinfo {author} {\bibnamefont {{A. P. Severyukhin}}},
  \bibinfo {author} {\bibnamefont {{S. \AA{}berg}}}, \bibinfo {author}
  {\bibnamefont {{N. N. Arsenyev}}}, \ and\ \bibinfo {author} {\bibnamefont
  {{R. G. Nazmitdinov}}},\ }\href@noop {} {\bibfield  {journal} {\bibinfo
  {journal} {Phys. Rev. C}\ }\textbf {\bibinfo {volume} {98}},\ \bibinfo
  {pages} {044319} (\bibinfo {year} {2018})}\BibitemShut {NoStop}%
\bibitem [{\citenamefont {Andreozzi}\ \emph {et~al.}(2007)\citenamefont
  {Andreozzi}, \citenamefont {Knapp}, \citenamefont {Lo~Iudice}, \citenamefont
  {Porrino},\ and\ \citenamefont {Kvasil}}]{Andreozzi_2007}%
  \BibitemOpen
  \bibfield  {author} {\bibinfo {author} {\bibfnamefont {F.}~\bibnamefont
  {Andreozzi}}, \bibinfo {author} {\bibfnamefont {F.}~\bibnamefont {Knapp}},
  \bibinfo {author} {\bibfnamefont {N.}~\bibnamefont {Lo~Iudice}}, \bibinfo
  {author} {\bibfnamefont {A.}~\bibnamefont {Porrino}}, \ and\ \bibinfo
  {author} {\bibfnamefont {J.}~\bibnamefont {Kvasil}},\ }\href@noop {}
  {\bibfield  {journal} {\bibinfo  {journal} {Phys. Rev. C}\ }\textbf {\bibinfo
  {volume} {75}},\ \bibinfo {pages} {044312} (\bibinfo {year}
  {2007})}\BibitemShut {NoStop}%
\bibitem [{\citenamefont {Bianco}\ \emph {et~al.}(2012)\citenamefont {Bianco},
  \citenamefont {Knapp}, \citenamefont {Lo~Iudice}, \citenamefont {Andreozzi},\
  and\ \citenamefont {Porrino}}]{Bianco_2012}%
  \BibitemOpen
  \bibfield  {author} {\bibinfo {author} {\bibfnamefont {D.}~\bibnamefont
  {Bianco}}, \bibinfo {author} {\bibfnamefont {F.}~\bibnamefont {Knapp}},
  \bibinfo {author} {\bibfnamefont {N.}~\bibnamefont {Lo~Iudice}}, \bibinfo
  {author} {\bibfnamefont {F.}~\bibnamefont {Andreozzi}}, \ and\ \bibinfo
  {author} {\bibfnamefont {A.}~\bibnamefont {Porrino}},\ }\href@noop {}
  {\bibfield  {journal} {\bibinfo  {journal} {Phys. Rev. C}\ }\textbf {\bibinfo
  {volume} {85}},\ \bibinfo {pages} {014313} (\bibinfo {year}
  {2012})}\BibitemShut {NoStop}%
\bibitem [{\citenamefont {De~Gregorio}\ \emph {et~al.}(2016)\citenamefont
  {De~Gregorio}, \citenamefont {Knapp}, \citenamefont {Lo~Iudice},\ and\
  \citenamefont {Vesel\'y}}]{De_Gregorio_2016}%
  \BibitemOpen
  \bibfield  {author} {\bibinfo {author} {\bibfnamefont {G.}~\bibnamefont
  {De~Gregorio}}, \bibinfo {author} {\bibfnamefont {F.}~\bibnamefont {Knapp}},
  \bibinfo {author} {\bibfnamefont {N.}~\bibnamefont {Lo~Iudice}}, \ and\
  \bibinfo {author} {\bibfnamefont {P.}~\bibnamefont {Vesel\'y}},\ }\href@noop
  {} {\bibfield  {journal} {\bibinfo  {journal} {Phys. Rev. C}\ }\textbf
  {\bibinfo {volume} {93}},\ \bibinfo {pages} {044314} (\bibinfo {year}
  {2016})}\BibitemShut {NoStop}%
\bibitem [{\citenamefont {Litvinova}(2015)}]{Litvinova_2015}%
  \BibitemOpen
  \bibfield  {author} {\bibinfo {author} {\bibfnamefont {E.}~\bibnamefont
  {Litvinova}},\ }\href {\doibase 10.1103/PhysRevC.91.034332} {\bibfield
  {journal} {\bibinfo  {journal} {Phys. Rev. C}\ }\textbf {\bibinfo {volume}
  {91}},\ \bibinfo {pages} {034332} (\bibinfo {year} {2015})}\BibitemShut
  {NoStop}%
\bibitem [{\citenamefont {Tselyaev}\ \emph {et~al.}(2020)\citenamefont
  {Tselyaev}, \citenamefont {Lyutorovich}, \citenamefont {Speth},\ and\
  \citenamefont {Reinhard}}]{Tselyaev_2020}%
  \BibitemOpen
  \bibfield  {author} {\bibinfo {author} {\bibfnamefont {V.}~\bibnamefont
  {Tselyaev}}, \bibinfo {author} {\bibfnamefont {N.}~\bibnamefont
  {Lyutorovich}}, \bibinfo {author} {\bibfnamefont {J.}~\bibnamefont {Speth}},
  \ and\ \bibinfo {author} {\bibfnamefont {P.-G.}\ \bibnamefont {Reinhard}},\
  }\href {\doibase 10.1103/PhysRevC.102.064319} {\bibfield  {journal} {\bibinfo
   {journal} {Phys. Rev. C}\ }\textbf {\bibinfo {volume} {102}},\ \bibinfo
  {pages} {064319} (\bibinfo {year} {2020})}\BibitemShut {NoStop}%
\bibitem [{\citenamefont {Lyutorovich}\ \emph {et~al.}(2016)\citenamefont
  {Lyutorovich}, \citenamefont {Tselyaev}, \citenamefont {Speth}, \citenamefont
  {Krewald},\ and\ \citenamefont {Reinhard}}]{Lyutorovich_2016}%
  \BibitemOpen
  \bibfield  {author} {\bibinfo {author} {\bibfnamefont {N.}~\bibnamefont
  {Lyutorovich}}, \bibinfo {author} {\bibfnamefont {V.}~\bibnamefont
  {Tselyaev}}, \bibinfo {author} {\bibfnamefont {J.}~\bibnamefont {Speth}},
  \bibinfo {author} {\bibfnamefont {S.}~\bibnamefont {Krewald}}, \ and\
  \bibinfo {author} {\bibfnamefont {P.-G.}\ \bibnamefont {Reinhard}},\ }\href
  {ArXiv: http://arxiv.org/abs/1602.00862} {\bibfield  {journal} {\bibinfo
  {journal} {Phys. At. Nucl.}\ }\textbf {\bibinfo {volume} {79}},\ \bibinfo
  {pages} {868} (\bibinfo {year} {2016})}\BibitemShut {NoStop}%
\bibitem [{\citenamefont {Kl{\"{u}}pfel}\ \emph {et~al.}(2008)\citenamefont
  {Kl{\"{u}}pfel}, \citenamefont {Erler}, \citenamefont {Reinhard},\ and\
  \citenamefont {Maruhn}}]{Kluepfel_2008}%
  \BibitemOpen
  \bibfield  {author} {\bibinfo {author} {\bibfnamefont {P.}~\bibnamefont
  {Kl{\"{u}}pfel}}, \bibinfo {author} {\bibfnamefont {J.}~\bibnamefont
  {Erler}}, \bibinfo {author} {\bibfnamefont {P.~.~G.}\ \bibnamefont
  {Reinhard}}, \ and\ \bibinfo {author} {\bibfnamefont {J.~A.}\ \bibnamefont
  {Maruhn}},\ }\href {http://dx.doi.org/10.1140/epja/i2008-10633-3} {\bibfield
  {journal} {\bibinfo  {journal} {Eur. Phys. J. A}\ }\textbf {\bibinfo {volume}
  {37}},\ \bibinfo {pages} {343} (\bibinfo {year} {2008})},\ \bibinfo {note}
  {http://www.arxiv.org/abs/0804.340}\BibitemShut {NoStop}%
\bibitem [{\citenamefont {Martin}(2007)}]{NDS_2007}%
  \BibitemOpen
  \bibfield  {author} {\bibinfo {author} {\bibfnamefont {M.}~\bibnamefont
  {Martin}},\ }\href {\doibase http://dx.doi.org/10.1016/j.nds.2007.07.001}
  {\bibfield  {journal} {\bibinfo  {journal} {Nuclear Data Sheets}\ }\textbf
  {\bibinfo {volume} {108}},\ \bibinfo {pages} {1583 } (\bibinfo {year}
  {2007})}\BibitemShut {NoStop}%
\bibitem [{\citenamefont {Heusler}(2020)}]{Heusler_2020}%
  \BibitemOpen
  \bibfield  {author} {\bibinfo {author} {\bibfnamefont {A.}~\bibnamefont
  {Heusler}},\ }\href {\doibase 10.1088/1742-6596/1643/1/012137} {\bibfield
  {journal} {\bibinfo  {journal} {Journal of Physics: Conference Series}\
  }\textbf {\bibinfo {volume} {1643}},\ \bibinfo {pages} {012137} (\bibinfo
  {year} {2020})}\BibitemShut {NoStop}%
\bibitem [{\citenamefont {Basu}\ and\ \citenamefont
  {Mccutchan}(2020)}]{NDS_2020_A=90}%
  \BibitemOpen
  \bibfield  {author} {\bibinfo {author} {\bibfnamefont {S.}~\bibnamefont
  {Basu}}\ and\ \bibinfo {author} {\bibfnamefont {E.}~\bibnamefont
  {Mccutchan}},\ }\href {\doibase https://doi.org/10.1016/j.nds.2020.04.001}
  {\bibfield  {journal} {\bibinfo  {journal} {Nuclear Data Sheets}\ }\textbf
  {\bibinfo {volume} {165}},\ \bibinfo {pages} {1} (\bibinfo {year}
  {2020})}\BibitemShut {NoStop}%
\bibitem [{\citenamefont {Heusler}\ \emph {et~al.}(2016)\citenamefont
  {Heusler}, \citenamefont {Jolos}, \citenamefont {Faestermann}, \citenamefont
  {Hertenberger}, \citenamefont {Wirth},\ and\ \citenamefont {von
  Brentano}}]{Heusler_2016}%
  \BibitemOpen
  \bibfield  {author} {\bibinfo {author} {\bibfnamefont {A.}~\bibnamefont
  {Heusler}}, \bibinfo {author} {\bibfnamefont {R.~V.}\ \bibnamefont {Jolos}},
  \bibinfo {author} {\bibfnamefont {T.}~\bibnamefont {Faestermann}}, \bibinfo
  {author} {\bibfnamefont {R.}~\bibnamefont {Hertenberger}}, \bibinfo {author}
  {\bibfnamefont {H.-F.}\ \bibnamefont {Wirth}}, \ and\ \bibinfo {author}
  {\bibfnamefont {P.}~\bibnamefont {von Brentano}},\ }\href {\doibase
  10.1103/PhysRevC.93.054321} {\bibfield  {journal} {\bibinfo  {journal} {Phys.
  Rev. C}\ }\textbf {\bibinfo {volume} {93}},\ \bibinfo {pages} {054321}
  (\bibinfo {year} {2016})}\BibitemShut {NoStop}%
\end{thebibliography}%

\end{document}